\documentclass[useAMS,usenatbib,referee]{mn2e}
\usepackage{epsfig}
\usepackage{graphicx}
\usepackage{lscape}
\usepackage[usenames,dvips]{color}

\title[Non-thermal emissions from 
middle-aged pulsars]
{Non-thermal emissions from outer magnetospheric accelerators of 
middle-aged pulsars}
\author[J. Takata and H.-K. Chang]{J. Takata $^{1}$\thanks{E-mail:
takata@tiara.sinica.edu.tw}  and  H.-K. Chang$^{2}$$^,$$^{3}$\\
$^{1}$Institute of Astronomy and Astrophysics,  and
 Theoretical Institute for Advanced Research in Astrophysics,
Academia Sinica; and National Tsing Hua University,
Taipei Taiwan\\
$^{2}$Department of Physics, National
Tsing Hua University, Hsinchu, Taiwan \\
$^{3}$ Institute of Astronomy, National
Tsing Hua University, Hsinchu, Taiwan 
} 
\begin{document}

\date{}

\pagerange{\pageref{firstpage}--\pageref{lastpage}} \pubyear{2002}

\maketitle

\label{firstpage}

\begin{abstract}
We discuss $\gamma$-ray emissions from the outer gap accelerators of 
 middle-aged pulsars for part of the series of our studies.  
A two-dimensional 
electrodynamic model is used to 
 solve the  distribution of  accelerating electric
fields with electron and positron pair creation and  radiation processes 
in the magnetic meridional plane. We compute the curvature radiation and the
synchrotron radiation by solving the evolution of the Lorentz factor 
and the pitch angle. The calculated spectra are compared with 
observed phase-averaged spectra. We also use a three-dimensional
geometrical model to discuss the pulse profiles. We argue  that 
the outer gap of  middle-aged pulsars  occupies the whole 
region  between 
the last-open field lines and the critical magnetic field lines, which are 
perpendicular to the rotational axis at the light cylinder. We assume that 
there is no  outer gap
accelerator inside the light cylinder between 
the rotational axis and the 
critical magnetic field lines. For the Geminga pulsar,  we demonstrate 
 that the outward curvature radiation dominates in the spectrum
above 10~MeV, while the inward synchrotron radiation dominates 
below 10~MeV. We find that the computed spectrum 
is consistent with the observations in X-ray through $\gamma$-ray
bands. With the pulse morphology of the $\gamma$-ray emissions, 
we argue that the inclination angle and the viewing angle for the Geminga
pulsar  are  $\alpha\sim 50^{\circ}$ and $\xi\sim 90^{\circ}$,
respectively.  We also apply  our method to another four middle-aged
 radio pulsars, whose spin-down power and distance from the
 Earth  expect the  possibility of  detection of  
$\gamma$-ray emissions from those pulsars by Fermi telescope. Applying the
 inclination angle and the viewing angle inferred from radio polarization
 characteristics, the visibility of the $\gamma$-ray emissions 
from the outer gap is discussed. 
We show  that $\gamma$-ray emissions from 
 PSRs B0355+54, B1449-64 and B1929+10 will  probably be detectable  
by  Fermi telescope. For PSR B0740-28, on the
other hand,  $\gamma$-ray emissions will not be detected because the
$\gamma$-ray beam from the outer gap 
will be oriented in a different direction from the viewing angle.

\end{abstract}


\section{Introduction}
\label{intro}
The Energetic Gamma-ray Experiment Telescope (EGRET) aboard the 
\textit{Compton Gamma-ray observatory} detected the gamma-ray 
emissions from 
six pulsars (Thompson 2004),  
 five of which show  pulsed radio emissions. The sixth, 
Geminga pulsar, has known as a radio quiet $\gamma$-ray pulsar 
or with very weak radio emissions (Vats et al. 1999). 
 Because of the observed  high brightness 
temperature of radio emissions from  pulsars, the radio radiation 
is a coherent mechanism. On the other hand, the 
emission mechanisms in optical through $\gamma$-ray bands are
non-coherent and non-thermal processes, which  have been discussed with 
the polar cap  model, the slot gap model and the outer gap model in
the literature. 

The polar cap accelerator 
 has been considered  as the origin of radio emission from  pulsars
 (Ruderman \& Sutherland 1975). The configuration of the 
dipole magnetic field  in inner magnetosphere region 
provides a natural explanation of position-angle curve of  polarization of
 the radio pulse.  
For  $\gamma$-ray emissions, however, the polar cap model 
has to invoke both 
a small inclination angle and a small viewing angle measured from the 
rotational axis to  reproduce the observed wide separation
 of the two peaks appeared in the pulse profiles (Daugherty \& Harding 1996).
 
The slot gap model (Arons 1983) 
is an extension of the polar cap model. 
The  possibility of a high-altitude extension of the accelerator near  
 the last-open field lines was proposed. 
The recent slot gap model explains 
 the spectrum in optical to $\gamma$-ray bands for the Crab pulsar
 (Muslimov \& Harding 2003, 2004; Harding et al. 2008) 
and the pulse profiles in $\gamma$-ray bands for the 
Vela pulsar (Dyks \& Rudak 2003).

The outer gap accelerator model assumes a strong acceleration  
between the null charge surface of the Goldreich-Julian charge density
 and the light cylinder (Cheng, Ho \& Ruderman 1986a, b). 
The Goldreich-Julian charge density 
 is described by $\rho_{GJ}\sim-\Omega B_z/2\pi c$ with $\Omega$  
being the rotation frequency, $B_z$  the magnetic field component 
projected to the rotation axis, and $c$  the speed of light. 
The recent studies of the  outer gap model also explain the 
spectra and  pulse profiles measured by  EGRET  (Hirotani 2007; 
Tang et al. 2008; Takata et al. 2008).   
Although numerous efforts have been done to discriminate  emission 
models with $\gamma$-ray data, the origin of the non-thermal emission
remains inconclusive.

In addition to the $\gamma$-ray data of  EGRET, 
 emission properties in lower energy bands have also been
provided. For example, Harding et al. (2002) 
found that the pulse profile of the Vela pulsar in  RXTE  bands 
shows four or five peaks in a single period instead of the
double peak structure measured by EGRET. The detail polarization features 
of  optical
pulsed emissions from the Crab pulsar have been measured by Kanbach et
al (2005). INTEGRAL also measured the polarization of the $\gamma$-ray 
emissions from the Crab pulsar for the first time (Dean et al. 2008). 
In the future, furthermore, 
 the polarization of X-ray and soft $\gamma$-ray emissions 
from the Crab pulsar will probably be able to be measured by ongoing projects
 such as PoGo (Kataoka et al. 2005) and the Nuclear Compton Telescope
 (Chang et al. 2007). Both the \textit{Astro
-rivelatore Gamma a Immagini LEggero} (AGILE) 
and recently lunched \textit{Fermi Gamma-Ray Space Telescope}
(Fermi, hereafter) 
will measure more detail features of the 
 phase-resolved spectra and the pulse profiles  
in 1~MeV through  100~GeV bands. 
In particular, Fermi telescope is expected to measure  
pulsed $\gamma$-ray emissions from both many radio pulsars and from 
unidentified EGRET sources. A comprehensive
theoretical  study with observations in multi-wavelength
bands is desired and will be important to discriminate emission models. 

Takata et al. (2007) and Takata \& Chang (2007) 
 discussed non-thermal  emissions from 
the young pulsars, PSRs B0531+21  (Crab) and B0540-69, 
 with the outer gap model. 
In Takata et al. (2007), they demonstrated 
 that the outer gap model can explain observed polarization 
properties of the optical pulsed emissions from the Crab pulsar.
 Takata \& Chang (2007) extended their calculation of the polarization to 
higher energy band and predicted  the results of future observations. 
Also, they produced  observed  
 phase-resolved spectra in optical through $\gamma$-ray bands for the
Crab pulsar. They  predicted that $\gamma$-ray flux on the Earth 
 from PSR B0540-69 is about $10^{-12}\mathrm{erg/cm^2s}$, 
which may be a target for  Fermi observation.  

Takata et al. (2008) discussed  the emission process 
in optical through  $\gamma$-ray bands  
 for the Vela-like young pulsars. Solving the electrodynamics of 
the outer gap, they first pointed out that 
 inward emission by ingoing particles is  an important 
component below 10~MeV bands, while only outward emission
 contributes to the spectrum above 10~MeV as the traditional outer gap model 
has assumed. They produced the observed multi-peak structure of 
the pulse profile in X-ray bands of the Vela pulsar with the inward and the 
outward emissions. They  predicted that the pulse profile at 1-10~MeV 
has  four peaks in a single period, which can  be checked 
in  future observations. 

We  had studied the non-thermal emissions from 
the Crab-like and the Vela-like pulsars. 
In this paper,  therefore, we  discuss the non-thermal emissions
from the outer gap of 
the middle-aged pulsars for part of the  series of  our studies. 
  Using a two-dimensional electrodynamic  model, 
we will solve the accelerating electric field with electron and
positron  pair creation and  
radiation processes (section~\ref{model}). 
We will also discuss the X-ray and $\gamma$-ray pulse profiles using 
a three-dimensional geometrical model (section~\ref{geomod}).
We apply the outer gap model to the Geminga pulsar 
(section~\ref{Gemingapul}).
The calculated spectrum will be compared with the observed
phase-averaged  
spectrum in X-ray through $\gamma$-ray bands. (section~\ref{spectr}). 
We discuss the inclination angle, which is angle 
 between the rotational axis and the
magnetic axis,  and the viewing angle  with the observed  properties
of the $\gamma$-ray pulse profile  for the Geminga
pulsar. (section~\ref{lightr}).
 Also by assuming that the radio
 beam from the Geminga pulsar is oriented in a different direction from the 
 line of sight, we will further constrain  the magnetic inclination
angle (section~\ref{radioq}).

We will also apply the model to another middle-aged radio pulsars 
(section~\ref{other}), whose timing properties 
 are  similar to  the Geminga pulsar. We will show candidates
for detections by Fermi telescope. 
With its  high sensitivity, Fermi telescope will also be 
 able to identify  $\gamma$-ray quiet radio pulsars, although 
their  $\gamma$-ray emissions should be detectable  
in the sense  of magnitude of the spin-down luminosity and its distance. 
With the present outer gap study, we will argue that 
PSR B0740-28 is a candidate of the  $\gamma$-ray quiet pulsars due to
misalignment of the directions of the 
$\gamma$-ray beam  and  the line
of sight.

\section{Magnetospheres of middle-aged pulsars}
\label{magneto}
\begin{figure}
\begin{center}
\includegraphics[width=7cm, height=7cm]{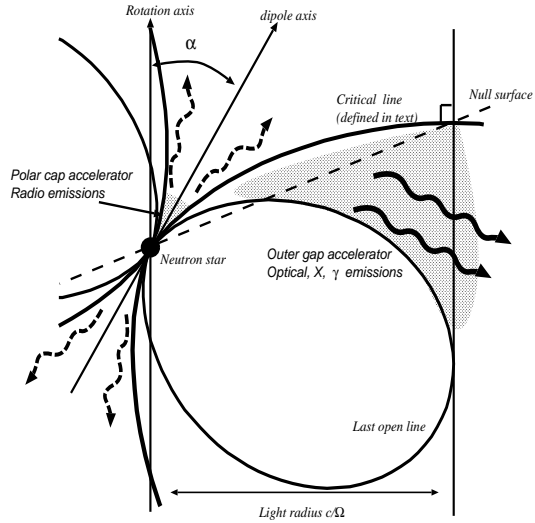}  
\caption{Schematic figure of  the magnetosphere of middle-aged pulsars.
The outer gap (filled region) 
occupies region between the last-open field line and 
the critical field line  
The radio wave originates from the inner region of the magnetosphere.
The optical to 
$\gamma$-ray photons 
are  from the outer gap region. The critical field line (thick solid line) 
is defined by the magnetic field line on which
 the Goldreich-Julian density vanishes at the light cylinder.} 
\label{pulsar}
\end{center}
\end{figure}

Figure~\ref{pulsar} shows a  schematic picture of the magnetosphere model 
for middle-aged pulsars. 
In general,  we  expect that the outer gap can arise between the 
last-open field lines and the magnetic field lines 
(so called critical field lines, hereafter) on which 
the Goldreich-Julian charge density vanishes at the light cylinder. 
Between the rotational axis and  the critical field lines, 
the null charge surface is located beyond the light cylinder so that 
the Goldreich-Julian charge density  does not change its sign 
 all the way from 
the stellar surface to the light cylinder. Therefore, we
 expect that there is no outer gap  in the region  
between the rotational axis and the critical field lines.  

For a static dipole field,  
the polar angle of  the critical field line on the stellar surface 
in the magnetic meridional plane
 is 
\begin{equation}
\theta_c=\alpha+\sin^{-1}\left[\sin(\theta_n-\alpha)
\left(\frac{R_*}{R_{lc}}\sin\theta_n\right)^{1/2}\right],
\label{crit}
\end{equation}
 where $\alpha$ is the magnetic inclination angle, $R_{*}$ is the stellar 
radius, $R_{lc}=c/\Omega$ is the light radius, and $\theta_n$ 
is the angle 
of the null charge surface of the Goldreich-Julian density;
\begin{equation}
\theta_n=\tan^{-1}\left(\frac{3\tan\alpha+\sqrt{8+9\tan^2\alpha}}{2}\right).
\end{equation}
We define the trans-field thickness of the maximally extending gap 
in the magnetic 
 meridional plane  with $\delta\theta_{max}=\theta_{pc}
-\theta_{c}$, where $\theta_{pc}$ is the polar cap angle. 
For the static dipole field, the  polar cap angle is expressed by 
\begin{equation}
\theta_{pc,0}=\alpha+\sin^{-1}\left[\sin(\theta_{lc}-\alpha)
\left(\frac{R_*}{R_{lc}}\sin\theta_{lc}\right)^{1/2}\right],
\label{rim}
\end{equation}
where
\[
\theta_{lc}=\tan^{-1}\left(\frac{-3-\sqrt{9+8\tan^2\alpha}}
{4\tan\alpha}\right).
\]

 For the middle-aged pulsars, 
we can expect that the thickness of the outer gap 
becomes to be comparable with the maximum thickness
 $\delta\theta_{max}$ as follows. 
The potential 
drop on the stellar surface between the critical magnetic field line and
the polar cap rim 
is 
\begin{equation}
V_{a}\sim 1.3\times 10^{13}P^{-3/2}\dot{P}^{1/2}_{-15}(1-\sin\theta^3_n)
~~\textrm{Volt}, 
\label{drop} 
\end{equation}
which becomes $V_{a}\sim 1.7\times 10^{14}$~Volt for the Geminga
pulsar. 
  
We can show that
 almost whole potential drop $V_{a}$ is required to explain the GeV
emissions from the Geminga pulsar.  For the Geminga pulsar, 
a spectral cut off  around 2~GeV was measured (Fierro et al. 1998).
 We estimate whole potential drop in the gap,  $V_{gap}$, from the 
cut off energy. We assume that $\gamma$-ray photons are emitted via the 
curvature radiation, and that  the particles that emit the cut off photons 
are  accelerated by  the whole potential drop along the magnetic field lines 
in the gap. The typical strength of the accelerating electric field is
estimated from  
 $E_{||}\sim V_{gap}/R_{lc}$. 
We assume the force balance between the acceleration and 
the radiation back reaction force to calculate a typical Lorentz factor 
of the accelerated particles; 
$\Gamma\sim (3R_{c}^2E_{||}/2 e)^{1/4}\sim (3R_{c}^2V_{gap}/2 e R_{lc})^{1/4}$,
 where $R_{c}$ is the curvature radius 
of the magnetic field line. Using the typical energy of the curvature 
photons $E_c\sim 3hc\Gamma^3/4\pi R_c$,
 the estimated potential drop in the outer gap is  
\begin{equation}
V_{gap}\sim 2.8\times 10^{14} \left(\frac{E_c}{2~\textrm{GeV}}\right)^{4/3}
 \left(\frac{\Omega}{26.5~\mathrm{s^{-1}}}\right)^{-1/3}~~\textrm{Volt}
\label{drop1}
\end{equation}
where we used $R_{c}=R_{lc}$. We find that 
the estimated magnitude of the  potential drop  is comparable 
with the available 
potential drop  defined by equation of (\ref{drop}) for the Geminga pulsar.
The observed $\gamma$-ray emissions from the Geminga pulsar imply that 
 most of the region 
between the last-open field lines and the critical field lines will be occupied
 by the outer gap  for the middle-aged pulsars.

For younger pulsars,  the outer gap accelerator occupies  
about 10~\% of the whole region between 
 the last-open field lines and the critical
field lines.  For example, the available potential drop defined by
equation of (\ref{drop}) for the Vela pulsar becomes $V_a\sim 2.5\times
10^{15}$~Volt.  The observed cut off energy around 3~GeV indicates the
potential drop in the gap is about $V_{gap}\sim
10^{14}$~Volt. Therefore, only about 10~\% of the available potential
drop is required to explain the $\gamma$-ray emissions from the Vela
pulsar. 

From the dynamic point of view, we can also argue
  that the outer gap of the middle-aged pulsars occupies 
 most of the  region between the last-open field line and the
critical field line. The
$\gamma$-rays emitted by the accelerated particles may convert into
electron and positron pairs by the pair creation process 
with a soft X-ray field 
from the stellar surface. 
 If typical energy of the curvature photons  have an energy 
such that the pair creation condition $E_{\gamma}E_X\ge(m_ec^2)^2$ is
satisfied, the outer gap can not develop further due to  screening effects  
of the pairs. In such a case, the position of the upper boundary of the gap 
is determined by the pair creation condition. For the Geminga pulsar, however, 
the pair creation in the outer magnetosphere is not efficient. 
The temperature of the stellar surface 
is about $T_s\sim 0.5$~MK (Kargaltsev et al. 2005),  so
 the threshold energy of the $\gamma$-rays for the pair creation is
$E_{\gamma}\sim (m_ec^2)^2/kT_s\sim 6~$GeV, where $k$ is the Boltzmann
constant. From equation of (\ref{drop1}), the potential drop of
$V_{gap}\sim 10^{15}$~Volt  in the gap
 is required to accelerate the particles to emit 6~GeV photons.
However it will be impossible to accelerate the particles 
because  the available potential drop in the outer gap  $V_a\sim
2\times 10^{14}$~Volt estimated from equation of (\ref{drop}) 
is smaller than the required  potential 
drop for the pair creation.  The particles can not be 
 accelerated in the outer gap to emit 6~GeV photons 
with the curvature radiation process. Therefore, we  expect that 
the outer gap of the middle-aged pulsars can develop to occupy the
region between  the last-open field lines and the critical field lines.
As we said in the  first paragraph in this section, 
 we do not expect that the outer gap accelerator arises beyond the
critical field lines, because 
the Goldreich-Julian charge density  does not change its sign 
 all the way from the stellar surface to the light cylinder 

We assume that the radio  emission originates from the polar cap
accelerator, while the emissions in optical through $\gamma$-ray bands 
originate from
 the outer gap accelerator. Electrodynamically speaking,  because of 
the different flow directions of the electric currents
  in the polar cap and in the outer gap, 
 the current conservation forces that both acceleration 
regions do not exit on the same magnetic field lines.
In this paper, we anticipate 
that the outer gap accelerator is extending from the last-open 
field line to the critical field lines, and 
the polar cap accelerator  extends in a region 
between the critical field lines
and the rotational axis. This assumption will be rejected 
if the radio emissions take place 
on the last-open field line. In fact, there are no  
observational evidences. For example, 
the phenomenological study with the polarization characteristics 
for the radio emissions (Dyks et al. 2004) 
indicates that the outer rim of  the radio beam is 
emitted on the magnetic fields located  well above the last-open field lines.

\section{Two-dimensional electrodynamic outer gap  model}
\label{model}
In this section, we briefly summarize our method of the
electrodynamic study of the outer gap model. A more detail 
description of our model can be found 
in the previous paper (Takata et al. 2004).
 We solve the
accelerating electric field with the radiation and the pair creation
processes  in the magnetic meridional plane, which includes 
the rotation axis and the magnetic axis. In the electrodynamic study, 
we use a static dipole field to solve the structure of the outer gap. 
 We consider a magnetized rotator in which an inclination angle $\alpha$ 
between the rotational axis and magnetic axis is smaller than $90^{\circ}$.  

\subsection{Electrodynamics}
\label{gapst}
The stationary electric potential, $\Phi_{nco}$, 
 for the accelerating field in the observer frame 
 is obtained from  (Shibata 1995; Mestel 1999)
\begin{equation}
\triangle\Phi_{nco}(\mathbf{r})=-4\pi[\rho(\mathbf{r})
-\rho_{GJ}(\mathbf{r})],
\label{poisson}
\end{equation}
where $\rho(\mathbf{r})$ is the space charge density, and
$\rho_{GJ}(\mathbf{r})$ is the Goldreich-Julian charge density, for which 
we use $\rho_{GJ}=-\Omega B_z/2\pi c$.

For the steady state, the continuity equations of the particles are written as 
\begin{equation}
 \mathbf{B}\cdot\nabla\left(\frac{v_{||}N_{\pm}(\mathbf{r})}{B}\right)=\pm
S(\mathbf{r}),
\label{basic2}
\end{equation}
where $v_{||}\sim c$ is the velocity along the field line, 
  $S(\mathbf{r})$ is the source term due to the pair creation process, 
 and  $N_+$ and $N_-$  denote
the number density of the outgoing (positrons) and ingoing (electrons) 
 particles, respectively. 
In the outer magnetosphere, $\gamma+X\rightarrow e^++e^-$ pair creation 
 process contributes to the source term
 $S(\mathbf{r})$. We simulate the pair creation 
process with a Monte Carlo method, which is explicitly described in 
Takata et al. (2004, 2006). 

To solve the Poisson equation (\ref{poisson}), we impose
 the boundary conditions on the four boundaries, which are called 
as inner, outer, upper and lower boundaries.  
The inner and outer boundaries are defined by the surfaces on which 
the accelerating electric field is vanishes, that is,  $E_{||}=0$. 
On the inner, upper and lower boundaries, 
 the accelerating potential is equal to zero, that is, $\Phi=0$. 
On the inner boundary, because 
both conditions $E_{||}=0$ and $\Phi=0$ are imposed,  
we can not know the position of the inner boundary in
advance. We solve the position of the inner boundary with the dynamics. 

The lower boundary is defined by the last-open field line. 
For the upper boundary, we define with  a magnetic field line.
For the outer boundary, we define it  with  a  curve line, which  is 
 perpendicular to the magnetic field lines in the  magnetic meridian plane.  
For such a case, the upper part of 
the our gap extends beyond the light cylinder, 
 if we put the outer boundary of the lower part of the gap  
close to (but inside ) the light cylinder. 
The present static Poisson equation of (\ref{poisson}) 
will not give the  correct solutions  for
 the electric field outside 
the light cylinder. As we will see later (Figure~\ref{Geminga1}), however,
  most $\gamma$-rays are  emitted from the   
lower part of the gap. The spectral features will be  less affected  
 by the structure of the  upper part of the gap. 
We expect that the results do not change  much  between 
the present approximated and  correct treatments. 
 To calculate the spectra, we take into account the emissions inside 
the light cylinder.

\subsection{Particle motion}
\label{emipro}
To calculate the synchrotron and curvature radiations, 
we solve an evolution of the particle's momentum using  
the electric field distribution in the outer gap obtained by 
 the method described in section~\ref{gapst}.  
The equation of motion for momenta of the parallel 
($P_{||}/m_e c=\sqrt{\Gamma^2-1}\cos\theta_{p}$) 
and perpendicular ($P_{\perp}/m_e c=\sqrt{\Gamma^2-1}\sin\theta_{p}$) 
to the magnetic field lines are, 
respectively, described as (Harding et al. 2005; Hirotani 2006)
\begin{equation}
\frac{dP_{||}}{dt}=eE_{||}-P_{sc}\cos\theta_p,
\label{paraeq}
\end{equation}
and \begin{equation}
\frac{dP_{\perp}}{dt}=-P_{sc}\sin\theta_p+\frac{c}{2B}\frac{dB}{ds}P_{\perp},
\label{perpeq}
\end{equation}
where $\theta_p$ is the pitch angle, $P_{sc}$ represents the radiation drag 
of the synchrotron and curvature radiation, and the second term on the right
 hand side on equation~(\ref{perpeq}) represents the adiabatic change 
along the dipole field line. The radiation drag, $P_{sc}$, of 
the synchrotron-curvature radiation is described by (Cheng \& Zhang 1996), 
\begin{equation}
P_{sc}=\frac{e^2c\Gamma^4Q_2}{12r_c}\left(1+\frac{7}{r_c^2Q_2^2}\right),
\end{equation}
where 
\begin{equation}
r_c=\frac{c^2}{(r_B+R_c)(c\cos\theta_p/R_c)^2+r_B\omega_B^2},
\end{equation}
\begin{equation}
Q^2_2=\frac{1}{r_B}\left(\frac{r_B^2+R_cr_B-3R_c^2}{R_c^3}\cos^4\theta_p
+\frac{3}{R_c}\cos^2\theta_p+\frac{1}{r_B}\sin^4\theta_p\right)
\end{equation}
\begin{equation}
r_B=\frac{\Gamma m_ec^2\sin\theta_p}{eB},~~~~\omega_B=\frac{eB}{\Gamma m_ec}.
\end{equation}
\subsection{Model parameters}
\label{modelp}

The model parameters are the magnetic inclination angle $\alpha$ 
and the current components $(j_1, j_2)$, where $j_1$ (or $j_2$) is the current
carried by the positrons (or electrons) coming into the gap through the inner 
(or outer )
boundary.  We also use the outer boundary as the model parameter. 
As the lower boundary of the gap, we define the last-open field line. 
For the upper boundary of  the middle age pulsars,  which we discuss
 in the paper, we use the critical field line defined by equation of 
(\ref{crit}). Setting the model
parameters, we  solve the position of the inner boundary, the
accelerating electric field, and the current component carried by
particles created in the gap.

 The created current, and the resultant calculated flux are determined by the 
value of the injected currents at the inner boundary $j_1$ and at the 
outer boundary  $j_2$. In fact, the magnitude of the created 
current is less affected by the current $j_1$ injected at the inner boundary, 
but is determined  by $j_2$ injected at the outer boundary,  
 because  most of pairs are created between the ingoing $\gamma$-rays and 
the outward propagating surface X-rays. In this paper, we choose the value 
of the component $j_2$ so that the created pairs produce the comparable 
$\gamma$-ray flux with the observations. 

The created current and the $\gamma$-ray flux 
do not monotonically increase with the 
 current component  $j_2$.  This is because an increase of the created
 current strengthens screening of 
the accelerating electric field. With a smaller electric field,
 the $\gamma$-ray photons with a lower energy are produced, and therefore
 a smaller amount of the pairs are created. Namely, 
 the increment of the created current  is
 determined under the competition between 
 the increase of $j_2$ and the screening of the electric field.
For example, if we assume the current $j_2=0.003$ in units of Goldreich-Julian 
value, we obtain the created current of $j_g\sim 0.3$ for the Geminga pulsar 
with $\alpha=60^{\circ}$. Then if we double  
 the value of $j_2$, we find from the numerical results that 
the increments of the created current and 
the total luminosity are  factors of  $\sim1.3$ and $\sim1.2$, respectively. 
We find  that  
the computed $\gamma$-ray flux is not sensitive to the detailed
 values of the current components injected at the boundaries.

Since the magnetic field must be modified by the rotational 
and the plasma effects the vicinity of 
 the light cylinder (Muslimov \& Harding 2005), 
the last-open field line will be  also  different from the conventional 
magnetic surface,  which is tangent to the light cylinder for the vacuum case. 
For example, Romani (1996) assumed the last-open lines 
with the field lines which have the polar angle 
$\theta_{pc}\sim1.4^{1/2}\theta_{pc,0}$ at the stellar surface, 
where $\theta_{pc,0}$ is the conventional 
polar cap angle (\ref{rim}) for static dipole field. In the studies of 
the force-free magnetosphere,
 $\theta_{pc}\sim 1.36^{1/2}\theta_{pc,0}$ was obtained by Contopoulos et al.
 (1999) and $\theta_{pc}\sim 1.27^{1/2}\theta_{pc,0}$ by Gruzinov (2005). 
In the paper, we use $\theta_{pc}\sim 1.36^{1/2}\theta_{pc,0}$ for the 
last-open field lines. The position  of the last-open field 
lines affects the 
accelerating electric field, spectrum and the pulse profile. 
For example, 
if we adopt a vacuum solution,
the typical electric field in the gap is described 
as $E_{||}\sim B_*R_*^2\delta\theta_{gap}^2/cR_c$ with $B_*$ begin the 
stellar magnetic field (Cheng et al. 2000; Hirotani 2006). 
 Then as  the last-open field line is located on
 the magnetic
field line with a larger polar angle on the stellar surface,  we obtain a
stronger accelerating
electric field because (i) the trans-field thickness of the gap,
$\delta\theta=\theta_{pc}-\theta_c$, becomes large and (ii) the typical 
curvature radius of the magnetic field line, on which most $\gamma$-rays
are emitted, becomes small. As a result,  
 a spectrum of the radiation from the gap becomes hard.  
For the pulse profiles, the phase separation between two peaks becomes
wide as the  last-open field shift to inner magnetosphere from the
canonical field line, which is tangent to the light cylinder 
(Takata \& Chang 2007). 
 In fact, the calculated  $\gamma$-ray spectrum and pulse profiles with 
the last-open field lines for 
$\theta_{pc}\sim 1.36^{1/2}\theta_{pc,0}$ are 
more consistent with the observations for the Geminga pulsar 
than those for $\theta_{pc}\sim \theta_{pc,0}$ (section~\ref{result}).

\section{Three-dimensional geometrical model}
\label{geomod}
The observed pulse profile  provides an important  tool 
for diagnose the geometry of 
the emission region in the pulsar magnetospheres.
In particular,  number of the peaks in a single rotational period, 
 positions of the peaks in the pulse profile, 
and  morphology of the pulse profiles at different energy bands 
will  discriminate  the emission models. 

In  the framework of  the outer gap model (and also the slot gap model),
 the peaks appear  in the pulse profiles because 
 the photons emitted various point pileup due to the effect of the special 
relativity correction; the aberration of the emitting direction
 and the flight time (Romani \& Yadigaroglu 1995; Dyks \& Rudak 2003). 
To compute the pulse profile, we anticipate that the emission
 direction coincides with 
the particle motion. In the observer frame, 
 the particle motion along the field lines in 
 the north hemisphere is described by $\mathbf{v}=\pm v_0\mathbf{b}
+v_{co}\mathbf{e}_{\phi}$, where the plus (or minus) sign represents
 the outgoing (or ingoing) particles, $v_0$ is the velocity along 
the magnetic field line and is determined by the condition that 
$|\mathbf{v}|=c$, 
$\mathbf{b}$ is the unit vector of the magnetic field, for which 
we adopt a rotating dipole field in the observer 
frame,  $v_{co}$ is 
the co-rotating velocity, and $\mathbf{e}_{\phi}$ is the unit vector of 
the azimuthal direction.  The emission direction, 
$\mathbf{n}\equiv \mathbf{v}/c$, is 
interpreted in terms of the viewing angle $\xi=n_z$ and 
the pulse phase $\Phi=-\phi_e-\mathbf{n}\cdot \mathbf{r_e}/R_{lc}$, where 
$\phi_e$ and $r_e$ is the azimuthal emission direction and the radial distance 
 to the emission point, respectively. We use a rotating magnetic dipole 
field in the observer frame to compute the pulse profile.

The distribution of emissivity  
 affects the pulse profiles. 
For the distribution of the emissivity in the meridional plane, 
 we use the results of the 
two-dimensional electrodynamic model, which solves the trans-field 
distribution of the emissivity in
 the magnetic meridional plane (see section~\ref{lightr}). 
For the structure of the azimuthal direction, it is difficult to model 
without solving a three-dimensional dynamics. For the first order
 approximation, however, we may apply the function form of 
 the accelerating electric field for the vacuum solution. 
In the vacuum solution, the typical value of the electric field along a field 
line  is determined by $E_{||} (\phi)\propto R_n(\phi)^{-1}$ 
with $R_n$ being the curvature radius at the null charge surface and $\phi$ 
the magnetic azimuthal angle of the magnetic
 field on the stellar surface measured 
from the magnetic meridional plane
(Cheng el al. 1986a; Cheng et al. 2000). In such a case, 
the  number of the curvature photons emitted per second by a particle  
 and the typical energy of the emitted photons are  proportional to  
$N_c(\phi)\propto R_n^{-3/4}$ and $E_{\gamma}\propto R_n^{-1/4}$, 
respectively,  if we assume the force balance
 between the particle acceleration and the radiation drag. 
 For example, we obtain $R_n\sim 0.24R_{lc}$ for
 the magnetic meridional plane ($\phi=0$) and  
$R_n\sim 0.72R_{lc}$ for  $\phi=90^{\circ}$,  if we use
 $\theta_{pc}\sim 1.36^{1/2}\theta_{pc,0}$ for the last-open field 
lines with the inclination angle $\alpha=60^{\circ}$. 
 In this paper, therefore, when we compute the pulse profile, we assume 
that ratio of the photon number emitted on the  field lines of    
 $\phi=0^{\circ}$ and of $\phi=90^{\circ}$ is $[N(90^{\circ})/N(0^{\circ})]
=(0.72/0.24)^{-3/4}\sim0.42$.  
We do not consider the emissions from the magnetic
 field lines on which the radial distance to the null charge point 
is larger than $r=R_{lc}$.

\section{Results}
\label{result}

\subsection{The Geminga pulsar}
\label{Gemingapul}
In this section, we apply the model to the Geminga pulsar and
constrain the magnetic inclination and viewing angles. 
First, we will show structure of the outer gap that 
  occupies most of the region
between the last-open line and the critical field line 
(section~\ref{gapgeo}).
 Then we will show that the calculated spectra 
 are consistent with the observations from
X-ray through $\gamma$-ray bands (section~\ref{spectr}). 
We use the properties of the observed  $\gamma$-ray pulse profile to
constrain the magnetic inclination and viewing angles 
 of the observer (section~\ref{lightr}).
For the Geminga pulsar, the observed pulse profile in $\gamma$-ray bands 
shows double peak structure, and 
the phase separation of the two peaks is about 0.5 phase. Finally, we
further constrain the magnetic inclination angle 
by discussing visibility of the radio beam in section~\ref{radioq},
in which we will assume that the radio beam from the Geminga pulsar is
 oriented in a different direction from the  line of sight. For the
 Geminga pulsar, we find that the calculated spectrum with the stellar 
dipole momentum inferred from the rotating dipole model
 is too soft compared with the observation.  
For  the present calculations, 
we used double of the dipole momentum 
inferred from the rotating dipole model.

\begin{figure}
\begin{center}
\includegraphics[width=14cm, height=7cm]{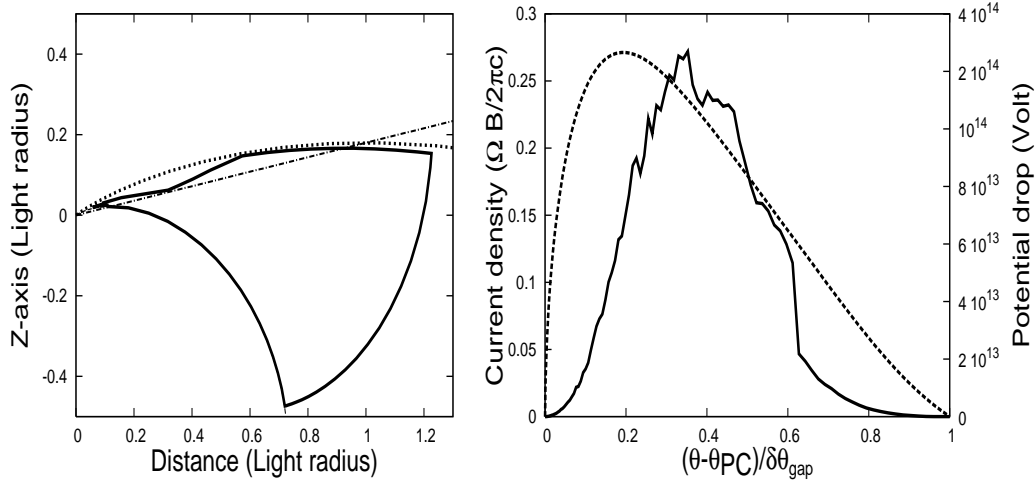}
\caption{Left; The geometry of the outer gap accelerator. The solid
line is the boundary of the gap. The dotted lines show 
 the critical line, and  the dashed-dotted line is 
the null surface of 
 the Goldreich-Julian charge density. 
 Right; The trans-field structure of the current (solid) and the electric 
potential drop (dashed line) along the magnetic field lines 
as a function of the polar angle of the field lines on the stellar
surface. We note that our static approximation
 will not be good treatment to describe the electric structure beyond
the light cylinder. But we expect that the gap thickness and 
the calculated spectrum will
not be affected much the electric structure beyond light cylinder, 
because most $\gamma$-ray photons are radiated from the lower
part outer gap inside the light cylinder. The results are for 
$\theta_{pc}=1.36^{1/2}\theta_{pc,0}$.}
\label{gapstr}
\end{center}
\end{figure}

\begin{figure}
\begin{center}

\includegraphics[width=10cm, height=7cm]{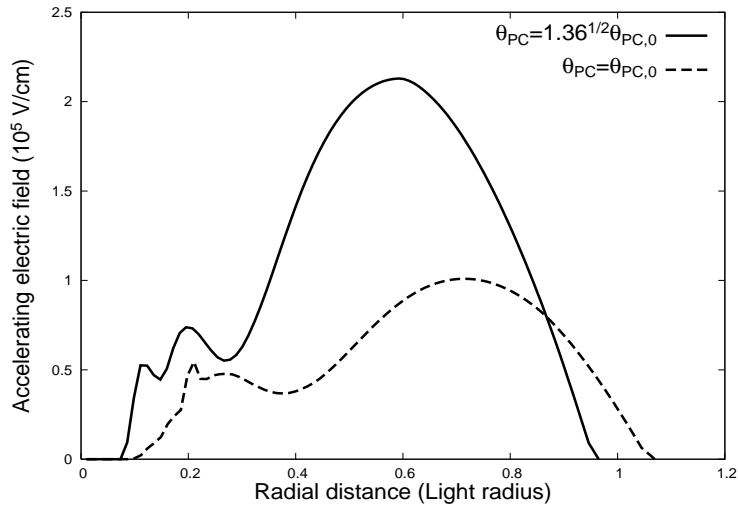}
\caption{The distribution of the accelerating electric field along 
the magnetic field line, on which the strength of
 electric field attains the maximum value. The solid and the dashed
 lines are results from the last-open field lines for 
$\theta_{pc}=1.36^{1/2}\theta_{pc,0}$ and  
$\theta_{pc}=1.36^{1/2}\theta_{pc,0}$, respectively.}
\label{ediftri}
\end{center}
\end{figure}

\subsubsection{Outer gap structure}
\label{gapgeo}
 Figure~\ref{gapstr} summarizes the outer gap structure of
the Geminga pulsar for
 the inclination angle  that $\alpha=60^{\circ}$, and for
 $j_{1,2}=3\cdot10^{-3}$, that is 0.3~\% of
the Goldreich-Julian currents are injected at both  inner and 
outer boundaries.  For the last-open field line,  we 
use the magnetic field line that has the polar angle 
$\theta_{pc}\sim1.36^{1/2}\theta_{pc,0}$, as we discussed 
in section~\ref{modelp}

In the left panel in Figure~\ref{gapstr},  the thick solid line
shows the boundary of the obtained gap structure. 
The dotted and dashed-dotted line  represent 
the critical field line and  
the null charge surface of the Goldreich-Julian charge density, respectively. 
As the computed boundary of the gap shows, we can find the solution 
 that the outer gap can occupy  
whole region between the last-open field line and the critical field
line, as we discussed  in section~\ref{magneto}.

\begin{figure}
\begin{center}
\includegraphics[width=7cm, height=7cm]{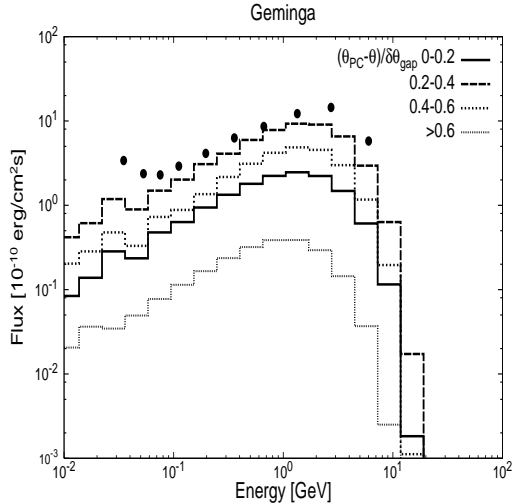}
\caption{The trans-field distribution of the emissivity of the $\gamma$-ray 
emissions. Each line represents the spectrum from the height
below 20~\% of the thickness measured
from the last-open field line, 
(solid line), between 20~\% and 40~\% 
(dashed line), between 40~\% and 60~\% (dotted line), and
above 60~\% (dashed-dotted line).}
\label{Geminga1}
\end{center}
\end{figure}

With the right panel of Figure~\ref{gapstr}, we display
the current density carried by the created pairs (solid line) and 
the total electric potential drop in the gap (dashed line), that is
potential drop between the inner and outer boundaries,  as a function of 
the polar angle, $\theta$, of the magnetic field lines on the stellar surface. 
In the abscissa, the polar angle satisfying 
the condition that $(\theta_{pc}-\theta)/\delta\theta_{gap}=0$ 
is corresponding to 
 the last-open field line, and the angle of
 $(\theta_{pc}-\theta)/\delta\theta_{gap}=1$ 
is corresponding to the upper boundary of the gap. 
From the solid line, we can see that 
 the maximum current density of about 30~\% of the 
Goldreich-Julian value runs through along the magnetic field line
 that penetrates the gap  
at  the height of about 40\% of the gap thickness measured from the last-open 
field line. From the dashed line, we also find that 
 the  largest  electronic potential drop of $V\sim 2\times 10^{14}$~Volt
 appears on the magnetic field line penetrating the gap at the height
of about 20\%  of the  thickness. This indicates that the strength 
of the  accelerating 
electric field becomes maximum on the  magnetic field 
line penetrating the gap at the height
of about 20\%  of the  thickness.  Because the emitting power of each 
magnetic flux tube is proportional to the product of  
the current density and the potential drop at each field line,
 we can read that most of emissions come from 
the height between 20~\% and 40~\% of the gap thickness
 (see also Figure~\ref{Geminga1}). 

Figure~\ref{ediftri} shows 
the distribution of the electric field along 
the magnetic field line, on which the strength of the electric field attains
 the maximum value in the outer gap.  We compare the 
results for $\theta_{pc}=1.36^{1/2}\theta_{pc,0}$ (solid line) and 
 $\theta_{pc}=\theta_{pc,0}$ (dashed line). As the Figure~\ref{ediftri}
 shows, a larger accelerating electric field is obtained if we assume 
 a larger polar cap angle. The typical strength of the accelerating 
electric field  is proportional to $E_{||}\propto
 (\delta\theta_{gap})^2/R_c$ (Cheng et al 2000; Hirotani 2006). In the present 
case,  the curvature radius  
on the field line on which the electric field attains the maximum value is
 typically $R_c\sim 0.8R_{lc}$ for $\theta_{pc}=1.36^{1/2}\theta_{pc,0}$ and 
$R_c\sim 1.2R_{lc}$ for  $\theta_{pc}=\theta_{pc,0}$.  This means that 
the ratio of strengths of the accelerating electric field for two cases 
is $E_{||}|_{\theta_{pc}=1.36^{1/2}\theta_{pc,0}}/
E_{||}|_{\theta_{pc}=\theta_{pc,0}}\sim 2$, which describes the difference 
of the strengths of the electric field seen in Figure~\ref{ediftri}.
 The typical energy of the photons emitted by the curvature radiation is 
proportional to $E_c\propto E_{||}^{3/4}R_c^{1/2}$. The ratio of 
the typical energies becomes $E_{c}|_{\theta_{pc}
=1.36^{1/2}\theta_{pc,0}}/E_{c}|_{\theta_{pc}=\theta_{pc,0}}\sim
1.4$. As a result, the calculated spectrum for 
$\theta_{pc}=1.36^{1/2}\theta_{pc,0}$ is harder than that for 
$\theta_{pc}=\theta_{pc,0}$.

\subsubsection{Spectrum} 
\label{spectr}
\begin{figure}
\begin{center}
\includegraphics[width=7cm, height=7cm]{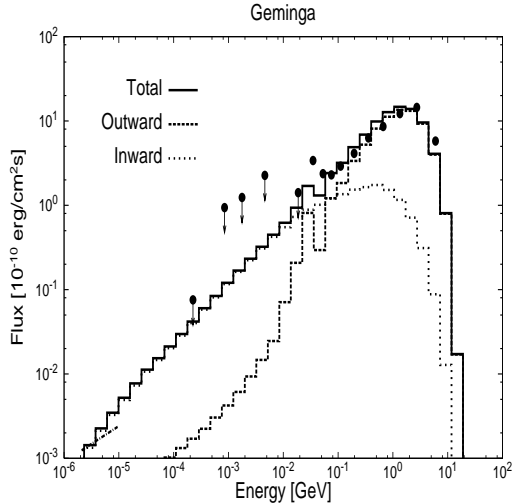}
\caption{The curvature-synchrotron spectrum for the
 Geminga pulsar in X-ray and $\gamma$-ray bands. 
The solid line represents the total emissions which includes the outward 
(dashed line) and the inward (dotted lines) emissions. The filled circles and 
the dashed-dotted line 
represent the observed phase-averaged spectrum (after Kargaltsev et al. 2005).}
\label{Geminga}
\end{center}
\end{figure}
Figure~\ref{Geminga} compares  
 the calculated spectrum of the curvature and 
synchrotron processes with the observations 
 in  X-ray through  $\gamma$-ray bands. The result is for the 
last-open field lines with the polar angle 
$\theta_{pc}=1.36^{1/2}\theta_{pc,0}$,
 and the model parameters are the same with that of 
Figure~\ref{gapstr}.
In Figure~\ref{Geminga},  
the filled circles and the dashed-dotted line 
show the observed phase-averaged 
spectrum. The dashed line and the
dotted line represent the emissions of  outgoing positrons and of the 
ingoing electrons, respectively. The solid line represents 
the total emissions. 
For each line, the components of the curvature and the synchrotron
 radiations are combined. 
In fact, the curvature radiation dominates in the spectrum above 100~MeV, 
while the synchrotron radiation dominates below 10~MeV. 
From Figure~\ref{Geminga}, we see that the total spectrum (solid line) is 
consistent 
with the observations in X-ray through $\gamma$-ray bands.  

If we use 
the canonical last-open field line with 
the polar angle $\theta_{pc}=\theta_{pc,0}$, the 
spectral cut off is lactated at the  energy factor of ~1/1.4 smaller 
than that  seen in the solid line of 
Figure~\ref{Geminga}, 
as we discussed in  the last paragraph of
 section~\ref{gapgeo}. Therefore, the calculated spectrum for the
 last-open 
field line with $\theta_{pc}=1.36^{1/2}\theta_{pc,0}$ is more consistent
 with that for the canonical last-open field line.

From Figure~\ref{Geminga}, we can see
 that the curvature radiation of 
outgoing positrons (dashed line) dominate in the total emissions above 
100~MeV.  This is because most of the pairs are created 
the vicinity of  the inner boundary by the inward propagating
 $\gamma$-rays and the outward propagating surface X-rays.
 The created  electrons and 
 positrons are accelerated 
inwardly and  outwardly by the electric field, respectively. 
The positrons created around the inner boundary are accelerated by almost 
whole potential drop in the gap, while the created electrons are accelerated 
by the potential drop only   
between the inner boundary and the point where they were created. Therefore, 
the total radiation power is larger for the outward emissions 
than the inward emissions. 
Below 10~MeV, on the other hand, we find that the synchrotron radiation 
of the ingoing electrons 
dominate in the calculated spectrum. This is because the synchrotron
emissions with strong magnetic field near  the stellar surface 
 contribute to the spectrum in X-ray bands.

\subsubsection{$\gamma$-ray pulse profiles}
\label{lightr}
\begin{figure}
\begin{center}
\includegraphics[width=14cm, height=14cm]{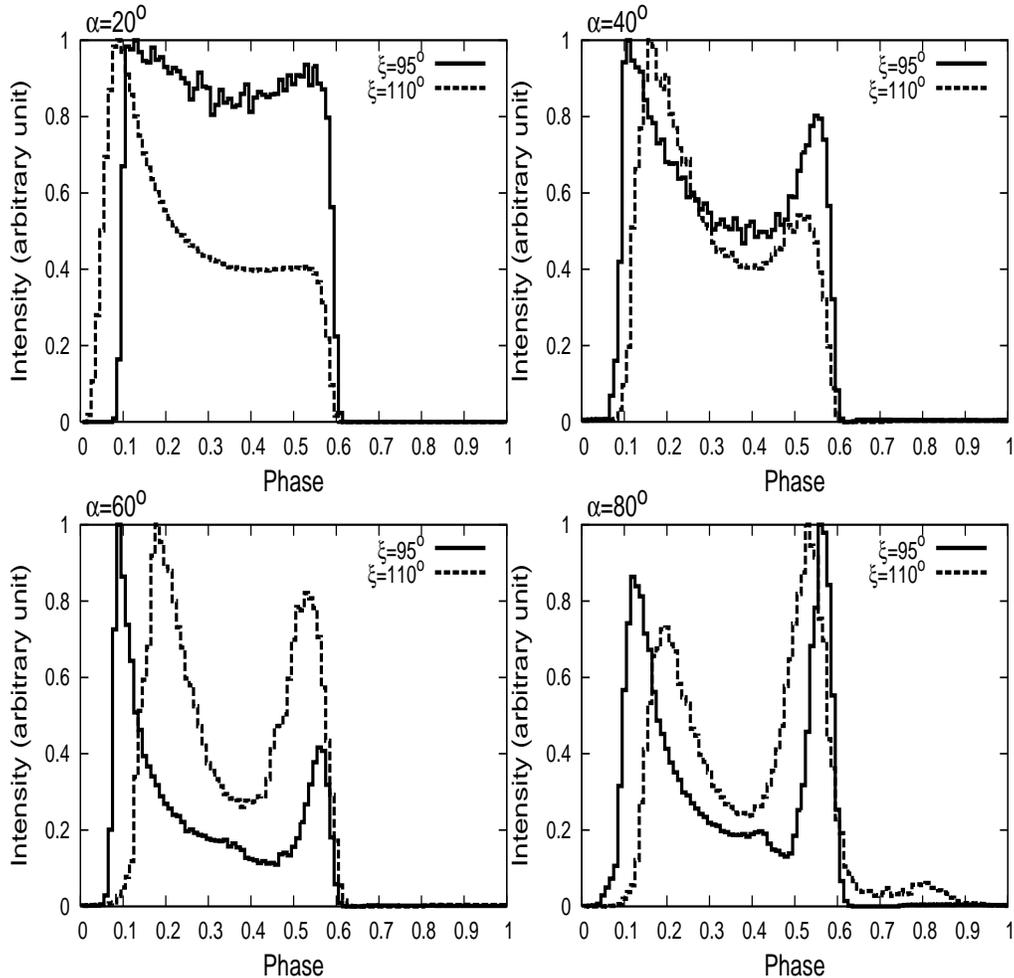}
\caption{Expected pulse profiles in the $\gamma$-ray bands for the Geminga
 pulsar. The upper left, the upper right, the lower left, 
and the lower right panels show the pulse profiles for the inclination angle 
of $\alpha=20^{\circ}$, $40^{\circ}$, $60^{\circ}$ and $80^{\circ}$. 
The solid and dashed lines in each pane represent the pulse profiles of the 
viewing angle  $\xi=85^{\circ}$ (or $95^{\circ}$) and $70^{\circ}$ 
(or $110^{\circ}$), respectively.
 The results are for $\theta_{pc}=1.36^{1/2}\theta_{pc,0}$.}
\label{light}
\end{center}
\end{figure}
\begin{figure}
\begin{center}
\includegraphics[width=7cm, height=7cm]{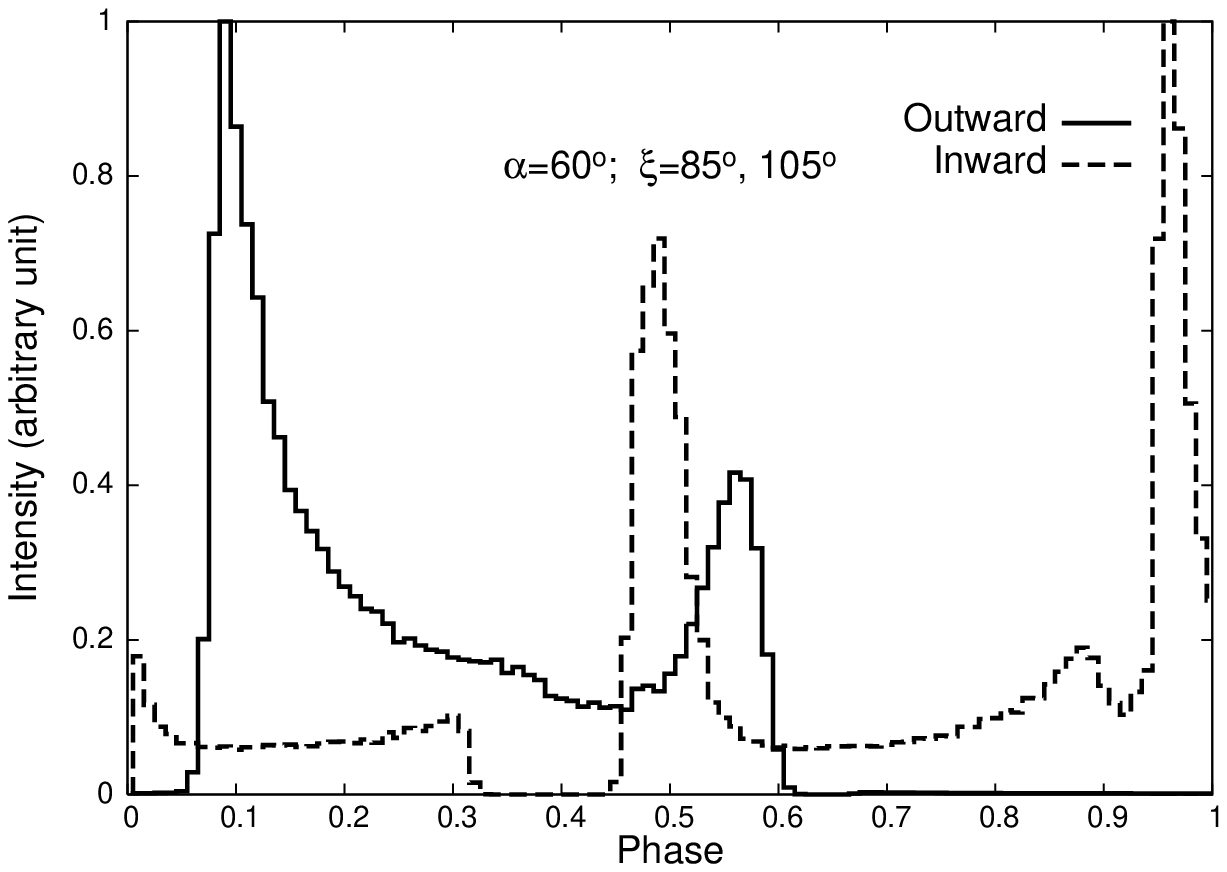}
\caption{
The pulse profile of the Geminga pulsar with the inclination 
angle $\alpha=60^{\circ}$ and the viewing angle $\xi=85^{\circ}$
 (or $105^{\circ}$). The solid and  dashed lines represent 
the pulse profiles for the outward and inward emissions, respectively.
 Above 100~MeV bands, only outward emissions contribute to the 
spectrum (Figure~\ref{Geminga}) so that the pulse profile (solid line) 
becomes double peak structure. 
For around 10~MeV bands, both the outward and inward emissions contribute 
to the pulsar profile, and the pulse profile has the  
four peaks in a single period. 
For X-ray bands, the inward emissions dominate in the pulse profile, which 
 has two peaks (dashed line) in a single period. 
}
\label{light2}
\end{center}
\end{figure}
We discuss the $\gamma$-ray pulse profiles, and constrain the 
inclination angle and the viewing angle using 
 the three-dimensional geometrical model described in section~\ref{geomod}. 
As Figure~\ref{Geminga} shows, the outward emissions dominate 
in the total spectrum above 100~MeV. In this section, therefore, 
we consider only the outward emissions to compute a pulse profile. 

 For the trans-field distribution of the emissivity 
 in the poloidal plane, we use the results of 
the two-dimensional electrodynamic study. In Figure~\ref{Geminga1}, 
we can read
 the trans-field distribution of the emissions above 10~MeV bands. 
 For example, the dashed line 
 represents the spectra of the emissions from the height 
 between 20~\% and 40~\% of the thickness measured
from the last-open field line. 
 By comparing the magnitude of the fluxes between the solid and dashed lines 
in Figure~\ref{Geminga1},  we may put an emissivity at  the height  
 below 20\% of the gap  about one order 
smaller than that at   the height 
 between 20\% and 40\% of the gap. 

Figure~\ref{light} summarizes the pulse profiles
 for  the magnetic inclination angle,
 $\alpha=20^{\circ}$ (upper left panel), 
  $\alpha=40^{\circ}$ (upper right panel),  $\alpha=60^{\circ}$ (lower
 left panel) and $\alpha=80^{\circ}$ (lower right panel).  
We calculate the emissions from the 
whole region between the last-open field lines 
and the critical field lines. 
 The results
 are for the last-open field lines with the polar angle 
$\theta_{pc}=1.36^{1/2}\theta_{pc}$. If we used 
the canonical last-open field lines with  
$\theta_{pc}=\theta_{pc}$ to  computed pulse profile,  we found that 
no geometries (inclination angle and viewing angle) can
 produce the observed phase separation between the two peaks with 
the $\gamma$-ray spectrum; we obtained narrower phase separation 
($\delta\phi<0.4$) than the observation $\delta\phi\sim 0.5$.

In Figure~\ref{light}, the solid and  dashed lines
 are results for the viewing angle of 
$\xi=85^{\circ}$ (or $95^{\circ}$) and $\xi=70^{\circ}$ (or $\xi=110^{\circ}$).
As the same with the conventional outer gap model,  
the first peak in the pulse profile is made by the emissions from 
 leading part 
of the gap according to the rotational direction, 
while the second peak is made by the emissions from trailing part of the
 gap. 
We can see from Figure~\ref{light} 
that the morphology of the pulse profiles depends on the inclination
 and  viewing angles. By
 comparing the observed pulse profile, therefore,  we can constrain
the  inclination  and viewing angles.
First we can rule out nearly aligned rotator, as represented
 by the pulse profile for $\alpha=20^{\circ}$. With such a small inclination 
angle, a sharp second peak does not appear 
in the pulse profile for any viewing 
angles of the observer. 
Secondary, we would rule out the viewing angle far 
away from $\xi=90^{\circ}$ as 
demonstrated by the pulse profile (dashed line) 
with $\xi=70^{\circ}$ in Figure~\ref{light}. 
In such a viewing angle, 
the pulse profile  
has a  single broad peak or the two peaks with a  narrower  separation of 
the phases compared with the observations ($\sim 0.5$~phase).
 Therefore, a larger inclination angle 
and the  viewing angle that  $\xi\sim90^{\circ}$, 
say $80^{\circ}<\xi<100^{\circ}$,   are preferred to 
explain the observed  pulse profile of $\gamma$-ray bands.

\subsubsection{Pulse profiles below 10~MeV}
In this section, we discuss the expected pulse profile in soft $\gamma$-ray
and  the X-ray bands. 
As the calculated spectrum in Figure~\ref{Geminga} shows, the inward emissions 
from the inward particles contribute to the total spectrum below 10~MeV.
Specifically, both the inward and the outward emissions contribute 
to the total emissions around 10~MeV, while the only  
inward emissions dominate into 
the total emission in X-ray bands. 
Therefore, we expect that the pulse profile below 10~MeV 
 is not simply an extension of that of the $\gamma$-ray bands above 100~MeV.
 Figure~\ref{light2} shows the pulse profile for the outward (solid) and 
the inward (dashed) emissions for $\alpha=60^{\circ}$ and 
$\xi=85^{\circ}$ (or $95^{\circ}$). 
  From the dashed line in Figure~\ref{light2},  
we find that the pulse profile of the inward emissions has also double 
peak structure. For the inward emissions, the first peak is made by 
the emissions from the trailing part of the gap and the  second peak 
is made by the emissions of the leading part of the gap. 

For the calculated spectrum around 10~MeV in Figure~\ref{Geminga}, both 
the outward and inward emissions contribute to the spectrum. 
The pulse profiles in 10~MeV bands will be made by  
both the  outward (solid line in Figure~\ref{light2}) and inward (dashed line 
in Figure~\ref{light2}) emissions. Therefore, we 
expect that the pulse profile  in soft $\gamma$-ray bands 
will have four peaks in a single period.

In X-ray bands, the inward emissions dominate in the total emissions 
(Figure~\ref{Geminga}). Therefore 
we expect the double peak structure of the pulse profile in the X-ray
bands for the Geminga pulsar.
 By comparing between the predicted pulse profiles in $\gamma$-ray 
(solid line) above 10~MeV bands  and X-ray 
bands (dashed line), 
the pulse peaks in X-ray band will appear in the 
leading phases of the pulses of the $\gamma$-ray bands.

With the present results, 
the observed differences  of the pulse  morphologies  in 
the different wavelength are strong tool to diagnose   
the emission processes in the pulsar magnetosphere.
For the Vela pulsar, the outer gap model can explain the morphology 
differences between the pulse profiles in RXTE and EGRET bands
(Takata et al.  2008). 
For the Geminga pulsar, the spectrum and  the pulse profiles observed 
by $ROAST$ and $ASCA$  indicated 
that the  non-thermal  X-rays and  $\gamma$-rays
 are produced by a distinct mechanism 
(Jackson et al. 2002).   Kargaltsev et al. (2005) analyzed X-ray spectrum and 
pulsations observed by $XMM-Newton$. They  found 
that the pulse profile of 
the non-thermal X-rays in 2-8~keV bands has  two peaks in a single 
period, and that 
the spectrum and  pulse profile  are  not simply an extension of that
 of the EGRET bands. The observed 
pulse morphologies for the Geminga pulsars 
are different in the different wave length. As we 
discussed above,  the present model 
 also predicts the double peak structure in X-ray pulse profile and 
the energy dependent pulse profiles from X-ray through $\gamma$-ray bands.
The results of the present model are generally consistent with the observed 
pulse profiles in X-ray and $\gamma$-ray bands. 
For the phase separation  between two peaks in $X$-ray bands, however, 
the observations in 2-8~keV bands imply about 0.35 (0.65)~phase 
 (Kargaltsev et al. 2005), while the present model 
predicts about  0.5~phase (dashed line in 
Figure~\ref{light2}). So, it may be  required an 
examination of the X-ray pulse profile
 with a more realistic magnetic field configuration.

\subsection{Radio ``quiet'' observations}
\label{radioq}
\begin{figure}
\begin{center}
\includegraphics[width=14cm, height=14cm]{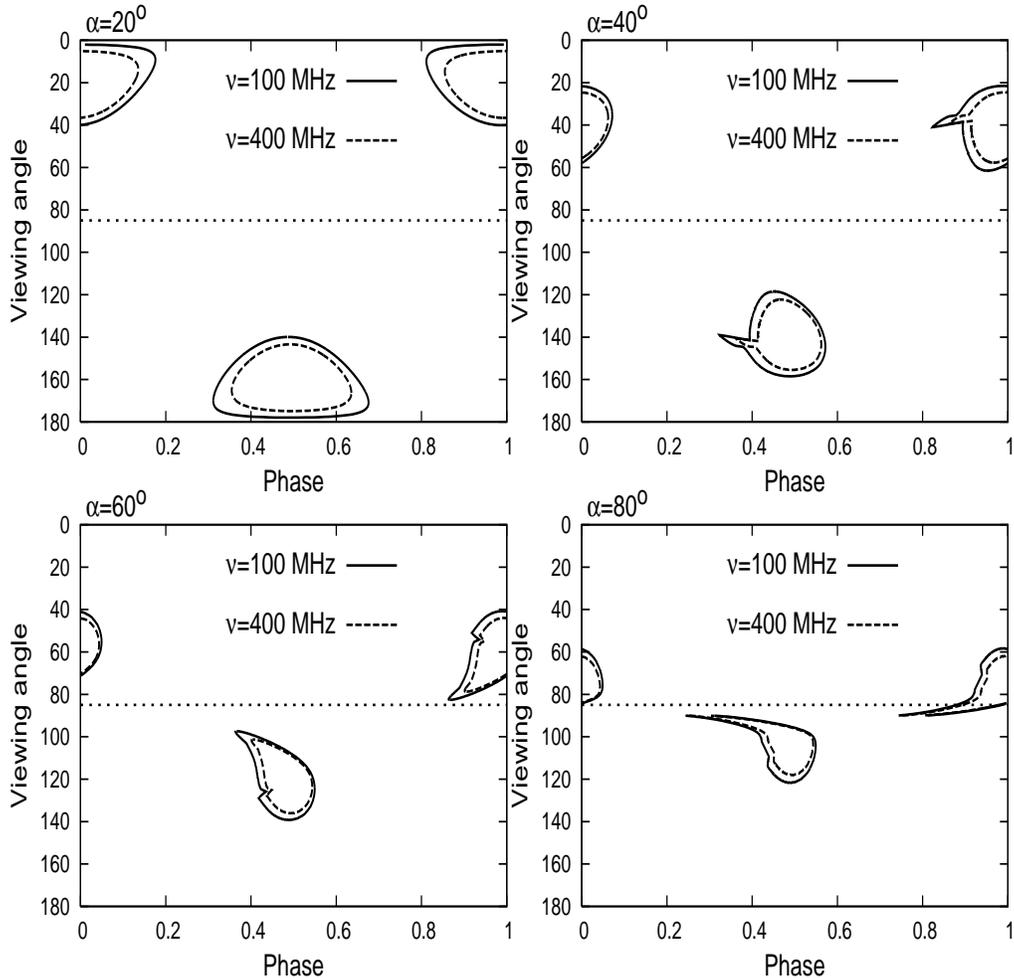}
\caption{Photon mapping of the outer rim of the radio beam emitted 
at the height described 
by equation of (\ref{height}) for different inclination angles 
$\alpha=20^{\circ}$ (upper left), $40^{\circ}$ (upper right),
 $60^{\circ}$ (lower left) and $80^{\circ}$ (lower right). The dotted line
 represent viewing angle of
$\alpha=85^{\circ}$. }
\label{radio}
\end{center}
\end{figure}

 The Geminga pulsar is the only known  
radio quiet (or  weak emissions) $\gamma$-ray pulsar. 
Based on the following reasons,
 we assume that  the radio beam of the Geminga pulsar  is oriented 
 in a   different direction from  the line of sight.
First, there are many radio pulsars around the Geminga pulsar on 
$P-\dot{P}$  diagram.
Second,  Lin \& Chang (2008) searched  candidate periods 
in unidentified EGRET sources for the next Geminga. 
They found that most of the candidate 
timing properties belong to that of 
Vela-like pulsars. If their candidate periods are true, their results 
 would  suggest that the major cause for a 
 radio quiet is likely geometrical rather than intrinsic.

The radio emission model with the 
 assumption that the radio ``quiet'' for the Geminga pulsar  is caused
by geometry  also provides a tool to study the magnetic inclination angle 
and the viewing angle with 
 the $\gamma$-ray emission model.
 For the emission height of the outer 
cone of the radio emissions, 
we apply  the empirical model developed by 
Kijak \& Gil (2003). They found a relation among the emission height, the 
rotation period, the period time derivative and the frequency of 
the radio wave.  Assuming the rim of the radio 
beam are emitted vicinity of  the critical magnetic field line (defined by 
$\theta_c$),  their relation is described as 
\begin{equation}
r\sim 40 R_{*}(\theta_{PC}/\theta_c)^2P^{0.3}(\dot{P}/10^{-15}
\mathrm{s s^{-1}})^{0.07}\nu^{-0.26}_{\mathrm{GHz}},
\label{height}
\end{equation}
where $R_{*}$ is the stellar radius and $\nu$ is the frequency of the 
radio wave.

Figure~\ref{radio} shows  photon mapping of the rim of the radio beam emitted 
at the height described 
by equation of (\ref{height}) for different inclination angles. 
We used the rotating dipole field 
as the magnetic field of the magnetosphere. 
In Figure~\ref{radio}, the solid and dashed lines represent the emissions 
 for the frequencies of $\nu=100~$MHz and 400~MHz, 
respectively. For the references the dotted line 
represents  line of sight for the viewing angle of $\xi=85^{\circ}$.  

For a small  inclination angle 
represented by  $\alpha=20^{\circ}$ (upper left panel in
Figure~\ref{radio}),   
 an observer with $\xi\sim 90^{\circ}$, which is preferred to explain 
observed $\gamma$-ray pulse profile, 
does not measure the radio emission. 
This is because  the magnetic pole points in   
a completely different direction from the line of sight for 
$\xi\sim90^{\circ}$. 
For largely inclined pulsar such as $\alpha=80^{\circ}$ 
(lower right panel), on the other hand, it is highly possible to be measured 
the radio emissions by the observer with $\xi\sim 90^{\circ}$, because 
the magnetic pole points close to the line of sight. For the Geminga pulsar, 
therefore, a smaller inclination angle 
 is preferred to explain the radio 
``quiet'' observations with the $\gamma$-ray detections.

Using the observed pulse profiles in  the $\gamma$-ray observations discussed 
in section~\ref{lightr},  
we safely eliminated the possibility of the nearly aligned rotator such as 
$\alpha=20^{\circ}$ because predicted single peak structure (upper left panel 
in Figure~\ref{light}) is not consistent with
 the observed double peak structure of 
the $\gamma$-ray pulse. 
Combing those  results obtained by analysis in  the $\gamma$-ray 
 and the radio ``quiet'' observations,  we conclude that the Geminga pulsar 
has a moderate  magnetic inclination angle 
of $\alpha\sim 50^{\circ}$, and we 
observe the Geminga pulsar with a viewing angle of $\xi\sim 90^{\circ}$. 

\subsection{Another middle-aged pulsars}
\label{other}
\begin{table}

\begin{tabular}{cccccccccc}
\hline
Pulsar & P  & $\dot{P}$ & $B$ & $\tau$ & $d$  & $\dot{E}/d^2$  & $\alpha$ & $\xi$ & Fermi  \\
 & (s) & ($10^{-13}$s/s) & ($10^{-12}$G) & ($10^5$yr) & (kpc)  & $10^{-10}\mathrm{erg/cm^2 s}$ &  &  &  \\
\hline\hline
Geminga & 0.2371 & 0.1097 & 1.63 & 3.42 & 0.16 & 1300 & $\sim
50^{\circ} $ & $\sim 90^{\circ}$ & O \\
\hline
B0355+54 & 0.1564 & 0.04397 & 0.839 & 5.64 & 1.10 & 38.9 & $51^{\circ}$ & $55^{\circ}$ & O \\
\hline
B0740-28 & 0.1668 & 0.1682 & 1.69 & 1.57 & 1.89 & 41.0 & $37^{\circ}$ & $27^{\circ}$ & $\times$ \\
\hline
B1449-64 & 0.1795 & 0.02746 & 0.71 & 10.4 & 1.84 & 5.88 & $ 55^{\circ}$ & $ 63^{\circ}$ & O \\
\hline
B1929+10 & 0.2265 & 0.01157 & 0.518 & 31 & 0.36 & 31.5 & $35^{\circ}$ & $ 60^{\circ}$ & O \\
\hline
\end{tabular}
\caption{Timing properties of the middle-aged pulsars. $P$ is the
rotational period, $\dot{P}$ is the time derivative of the rotational
period, $B$ is the stellar magnetic field inferred from the dipole
field model, $\tau$ is the pulse age, $d$ is the distance,
$\dot{E}/d^2$ is the spin-down luminosity measured on Earth, $\alpha$
is the magnetic inclination angle, and $\xi$ is the viewing angle. 
For the magnetic inclination angle $\alpha$ and the viewing angle
$\xi$,  we refer the present paper for the Geminga pulsar, Rankin
(1993) for PSRs B0355+54, B0740-28 and B1449-64, and Everett and
Weisberg (2001) for PSR B1929+10. In the last column, we show the
predicted detectability of pulsed $\gamma$-ray emissions by  Fermi telescope 
with $5\sigma$ in 50 hours.  }
\end{table}

We apply the present outer gap model to another middle-aged pulsars. 
Using ATNF 
Pulsar Catalogue (Manchester et al. 2005, 
see http://www.atnf.csiro.au/research/pulsar/psrcat), we chose four 
middle-aged 
pulsars,  PSRs B0355+54, B0740-28, B1449-64, and B1929+10, 
to discuss the detectability 
of the $\gamma$-ray emissions from the outer gap by 
 Fermi telescope. These four pulsars were
chosen with the following conditions. If 
the pulsars are at a large  distance 
from the Earth,   Fermi telescope could not measure the $\gamma$-ray 
emissions from the pulsars. A rough idea of the $\gamma$-ray flux on the 
Earth is described by 
$\eta L_{sp}/d^{2}$, where $\eta$ is efficiency, $L_{sp}$ is the 
spin-down luminosity, and $d$ is the distance to the pulsar. 
Our four pulsars have the spin-down luminosity so that  
$ \eta L_{sp}/d^2$ with $\eta=0.01$ 
of the Geminga pulsar is larger than  $10^{-12}~\mathrm{erg/cm^2 s}$, which 
is close to 
 the lower limit of  Fermi telescope detection with $5\sigma$ in 50 hours. 

Second, there is the  possibility that we can not measure the $\gamma$-ray
 emissions, because the $\gamma$-ray beam is oriented in a different 
direction from   the line of 
sight. Because the geometry of the $\gamma$-ray beams depends on 
the emission models, the detectability of the $\gamma$-ray emissions
 depends on the models. 
If we know the viewing angle and the magnetic inclination angle 
of the pulsar, we can test the models with  Fermi observations. 
On these ground,  we chose those four pulsars because the magnetic 
inclination  and  
viewing angles are inferred from the radio polarization data 
 (Rankin 1993; Everett \& Weisberg 2001). In Table~1, 
we summarize the timing properties of the four pulsars and the Geminga pulsar.
The represented results in this section were computed with (i) the 
 double of the dipole inferred from the rotating
 dipole model  and with (ii) the last-open field
 lines with the polar angle $\theta_{pc}=1.36^{1/2}\theta_{pc,0}$.

\subsubsection{PSR B0355+54}
\begin{figure}
\begin{center}
\includegraphics[width=7cm, height=7cm]{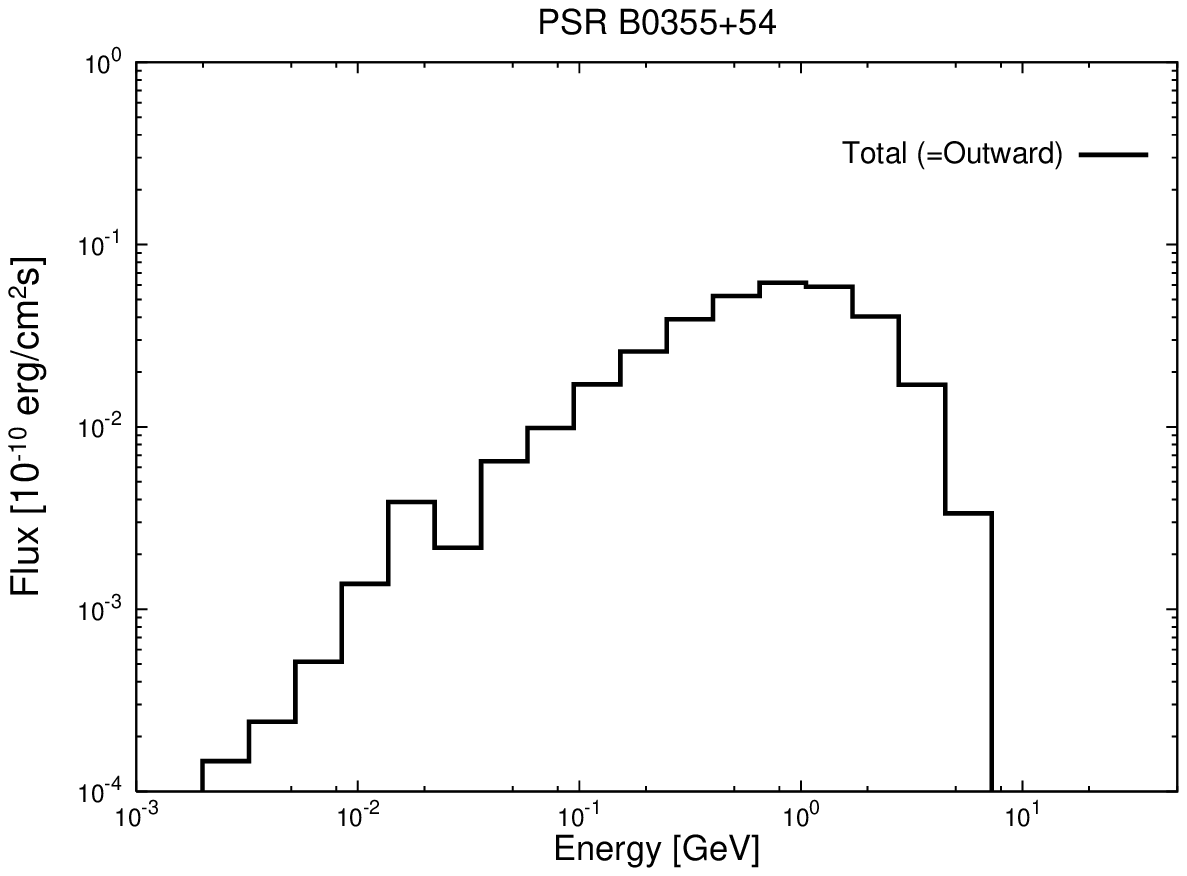}
\includegraphics[width=7cm, height=7cm]{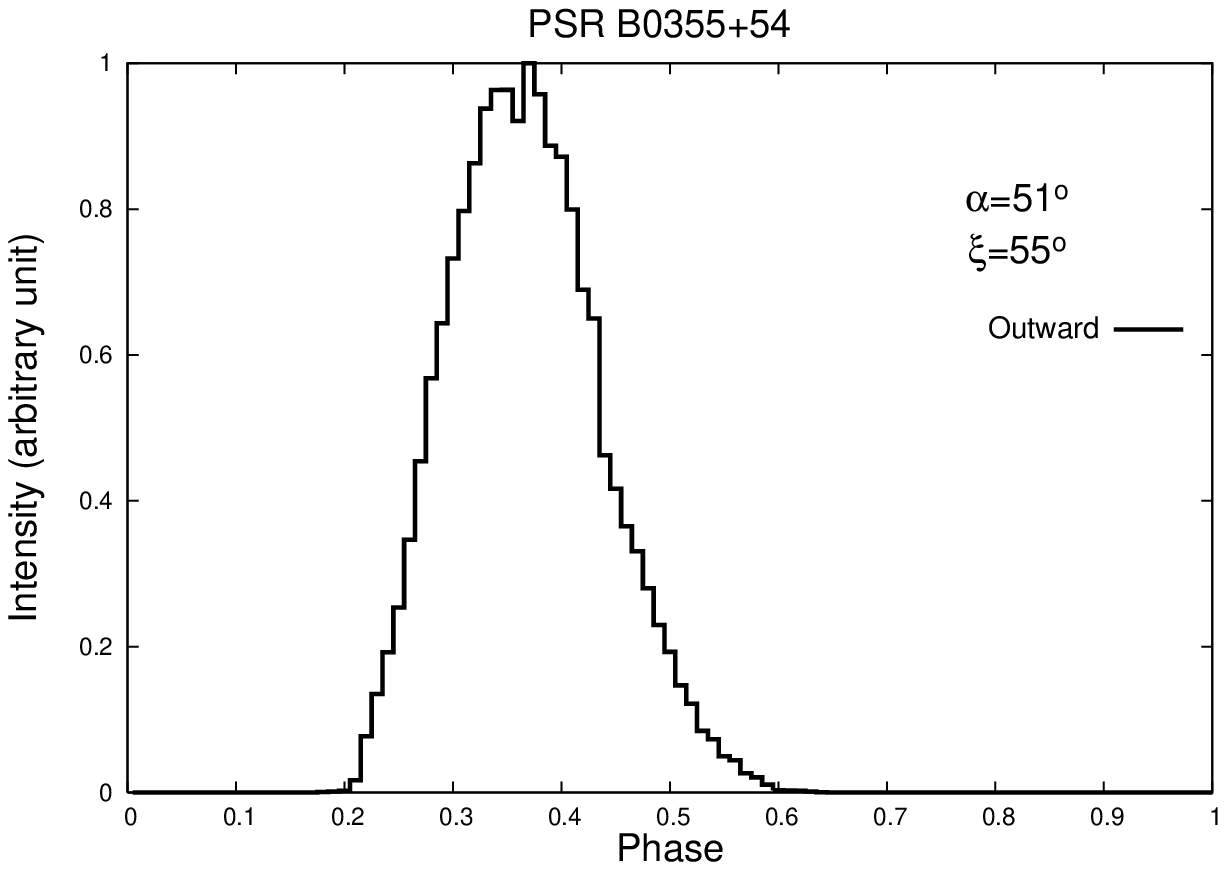}
\caption{The $\gamma$-ray spectrum (left) and pulse profile (right)
for PSR B0355+54. For  $\alpha=51^{\circ}$, 
the observe with  $\xi\sim 55^{\circ}$ measures only the outward emissions. }
\label{B0355}
\end{center}
\end{figure}
PSR B0355+54 has a rotational period  of $P=0.1564$~s 
(Manchester et al. 1972),  which is 
 the fastest in the four pulsars we chose. 
The distance  from the Earth is $d\sim1.1$~kpc 
(Hobbs et al. 2004). PSR B0355+54 is known to emit 
the X-ray (McGowan et al. 2006) and
 probably TeV $\gamma$-rays (Senecha et al. 1995) 
from its pulsar wind nebula.  The polarization data on the radio wave
 can be fitted by the rotational vector model; 
the magnetic inclination and the viewing angles are inferred as 
$\alpha\sim 51^{\circ}$ and $\xi\sim 55^{\circ}$, respectively (Rankin 1993).

Figure~\ref{B0355} shows the expected spectrum (left panel) 
and the pulse profile (right panel) in $\gamma$-ray bands for PSR B0355+54. 
For the current components injected at the inner and outer boundaries, 
we used $j_1=j_2\sim 0.01$. For the inclination angle $\alpha=51^{\circ}$, 
the observer with the  viewing angle $\xi=55^{\circ}$ can not measure 
the inward emissions from the outer gap.  
The model predicts the emission flux  around 1~GeV are 
about $F\sim 10^{-11}$ erg/s, which will be able to be measured by 
 Fermi telescope. With the  viewing angle $\xi\sim 55^{\circ}$, 
the expected pulse profile has a broad 
peak in a single rotational period, as right panel 
in Figure~\ref{B0355} shows. 

\subsubsection{PSR B0740-28}
PSR B0740-28 is younger and faster rotating pulsar than the Geminga pulsar 
(Bonsignori-Facondi et al 1973). The distance to the pulsar is $d=1.89$~kpc 
(Hobbs et al. 2004) and the magnitude  of  $L_{sp}/d^2$ 
is the largest in the four pulsars. The pulsar wind powered by PSR B0740-28 
was discovered in optical bands (Jones et al. 2002). The magnetic inclination 
angle $\alpha\sim 37^{\circ}$ and the viewing angle $\xi\sim 27^{\circ}$
 are inferred with the polarization characteristics of the radio wave 
(Rankin 1993).

We will not expect the detection of the $\gamma$-ray beam 
by  Fermi telescope  from PSR B0740-28, because  the $\gamma$-ray 
beam will be oriented in a different direction from  the line of sight. 
 Because the viewing angle with $\xi\sim 27^{\circ}$ points 
between the rotational and  magnetic axes in the magnetic
meridional plane, the observer measures mainly emissions inside the 
null charge surface and on the magnetic field lines toward 
rotational axis. However, the outer gap does not predict a strong
emissions from that region, but predicts a strong emission beyond 
the null charge surface and on the magnetic field lines curved away
from the rotational axis. The present outer gap model  expects 
that the strong emissions from the
outer gap of PSR B0740-28 
is oriented in a different direction from the line of sight.

For the slot gap model (Harding et al. 2008),
 on the other hand, a detection of 
 $\gamma$-ray emissions from 
PSR B0740-28 may be expected, because the model assumes the emissions from 
the polar cap region to the vicinity of  the light cylinder above the  whole 
last-open field lines (Dyks \& Rudak 2003). Therefore, 
PSR B0740-28 will be a good candidate to discriminate the outer gap model 
and the slot-gap model with  Fermi telescope.

\subsubsection{PSR B1449-64}
\begin{figure}
\begin{center}
\includegraphics[width=7cm, height=7cm]{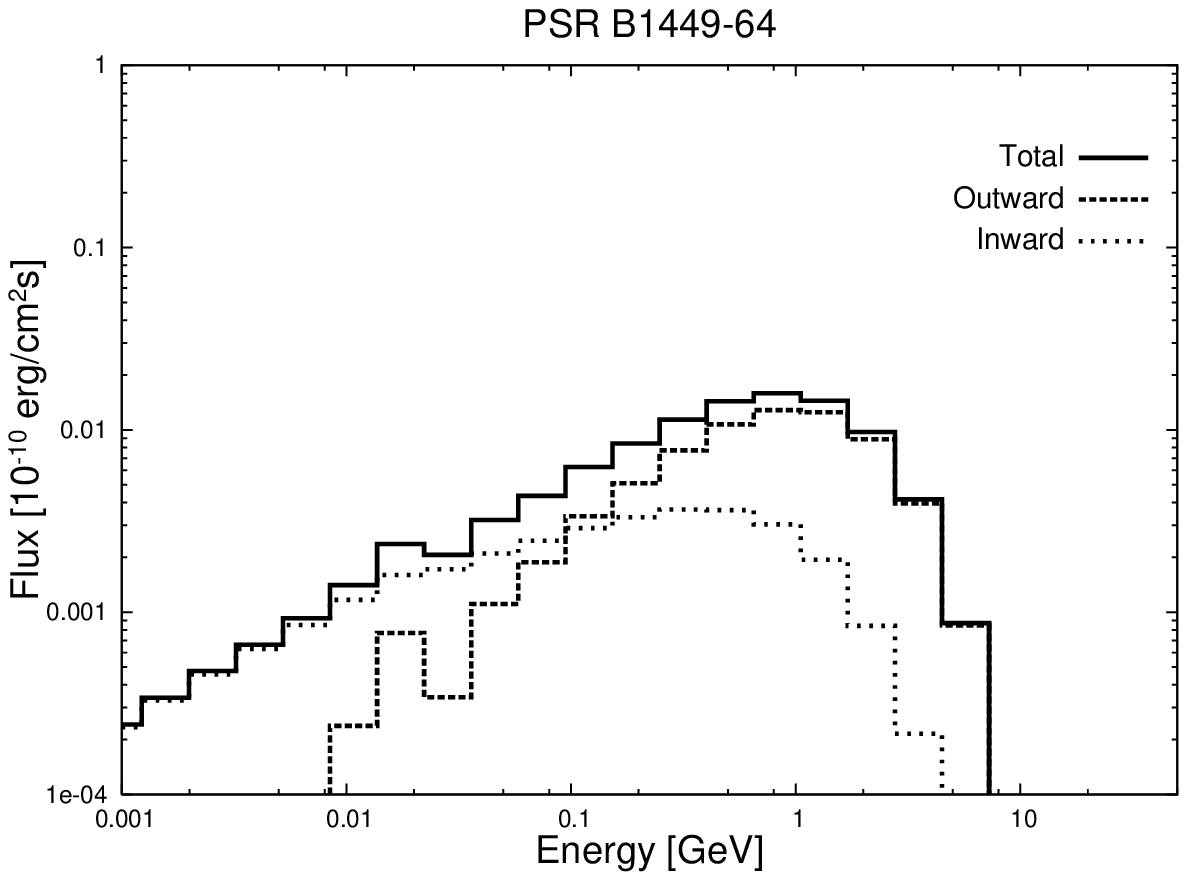}
\includegraphics[width=7cm, height=7cm]{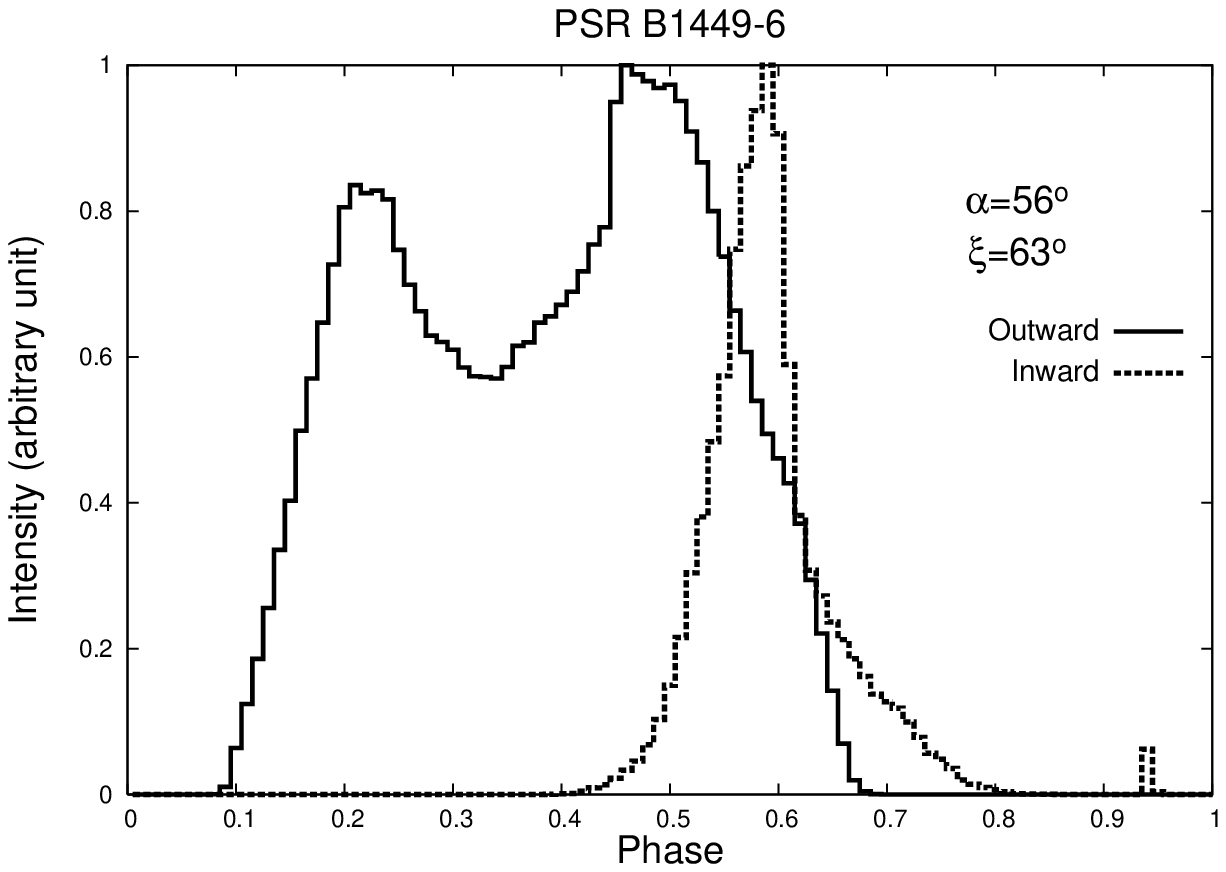}
\caption{The $\gamma$-ray spectrum (left) and pulse profile (right)
for PSR B1449-64.In the left panel, the solid, dashed and dotted lines 
represent the spectra of the total, outward and inward emissions.
 In the right panel, the solid and dashed  show the pulse profile of 
the outward and inward emissions, respectively.}
\label{B1449}
\end{center}
\end{figure}
PSR B1449-64  has  the rotational period 
$P= 0.1785$ and the period time derivative $\dot{P}=2.746\cdot 10^{-15}$~s/s 
(Large et al. 1969). The distance to the pulsar is $d=1.84$~kpc
 (Siegman et al. 1993). The magnetic inclination of $\alpha\sim 56^{\circ}$ 
and the viewing angles of $\xi\sim63^{\circ}$ are used to fit the polarization 
data of the radio wave with the rotating vector model (Rankin 1993). 
Figure~\ref{B1449} shows the expected spectrum (left panel) 
and the pulse profile (right panel) in $\gamma$-ray bands for PSR B1449-64. 
The expected pulse profile shows the double peak structure. 
We adopt $j_1=j_2=0.02$ for the current components injected at 
the inner and the outer boundaries.

\subsubsection{PSR B1929+10}
\begin{figure}
\begin{center}
\includegraphics[width=7cm, height=7cm]{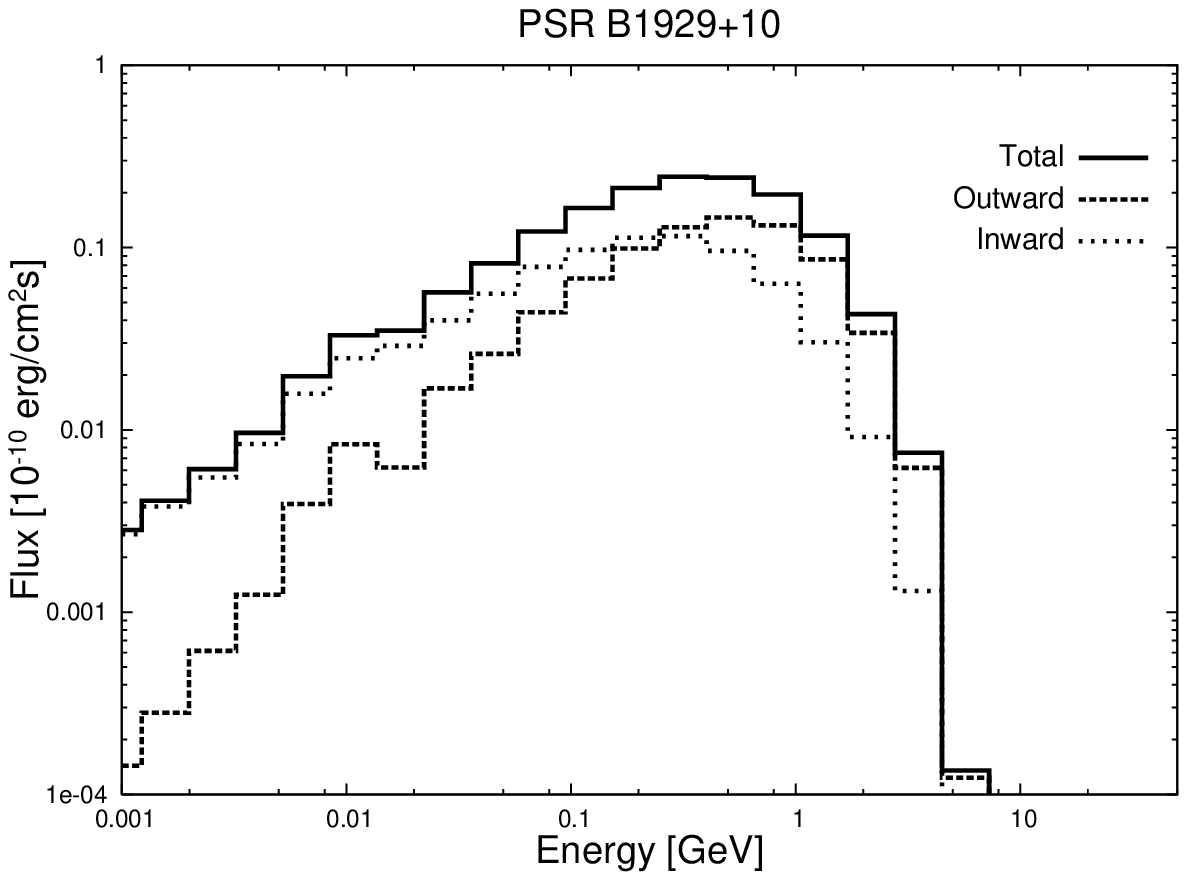}
\includegraphics[width=7cm, height=7cm]{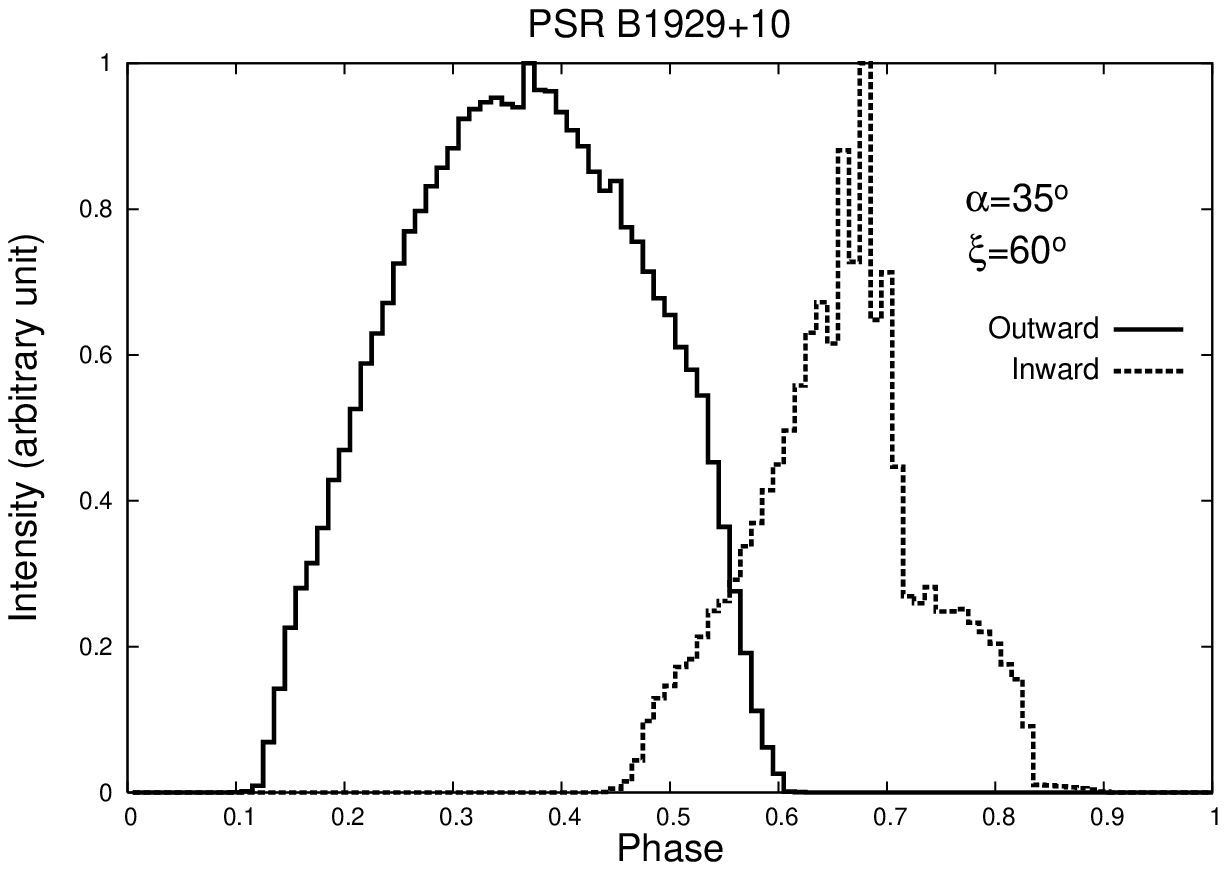}
\caption{The $\gamma$-ray spectrum (left) and pulse profile (right)
for PSR B1929+10.The lines correspond to same case as Figure~\ref{B1449}.}
\label{B1929}
\end{center}
\end{figure}
PSR B1929+10 is known as one of the closet pulsar
 with the distance $d\sim 0.36$~kpc (Hobbs et al. 2004). 
 PSR B1929+10 is the oldest pulsar $\tau\sim 3\cdot 10^{6}$~yrs 
in the pulsars we chose with its rotation period $P=0.2265$~s 
 and  the period time derivative $\dot{P}=1.157\cdot 10^{-15}$~s/s. 
The X-ray emissions from both the pulsar and the pulsar wind nebula have been 
reported (Becker et al. 2006; Hui and Becker 2008). 
The inferred  magnetic inclination and the viewing angle of the observer 
from the radio polarization are $\alpha\sim 35^{\circ}$ and 
$\xi\sim 60^{\circ}$, respectively (Everett and Weisberg 2001). We use 
$j_1=j_2=0.1$ for the currents injected the inner and outer boundaries.

Because  of the closeness to the Earth of PSR B1920+10,  
the expect $\gamma$-ray flux can significantly exceed 
the lower limit of 
 Fermi telescope detection with $5\sigma$ in 50 hours (Figure~\ref{B1929}), 
although the spin-down power and therefore the $\gamma$-ray luminosity are 
the smallest in the four pulsars. For the pulse profile, we expect 
 single peak in a single rotation period. 

\section{Summary}
We summarize the results of our work as follows. 

1.~We have studied the non-thermal emissions of the Geminga pulsar 
with the outer gap accelerator model. We solved the accelerating electric 
field with the pair creation and the radiation processes 
in the magnetic meridional plane. We demonstrated that  
the calculated spectrum of the curvature and the synchrotron emissions 
is consistent with the observation in X-ray through  $\gamma$-ray bands. 
The outward curvature radiation
 dominates in the spectrum of $\gamma$-ray bands,
 while the inward synchrotron emissions dominate in X-ray bands.  

2. We demonstrated that the $\gamma$-ray pulse above 100~MeV 
has two peaks, which are  made by the outward 
emissions. Around 10~MeV bands, we expect four peaks in a single 
period with both the outward and inward emissions. In X-ray bands, 
the pulse profile  has two peaks made by the inward emissions.
 The pulse phase in X-ray bands are not 
in phase with that in the $\gamma$-ray bands. 
The model  predicts the energy dependent
 pulse morphology in X-ray through $\gamma$-ray bands, which is 
also indicated by the observations. 
(Jackson et al. 2002; Kargaltsev et al. 2005).

3.~We discussed the inclination angle and 
the viewing angle for the Geminga pulsar
 with the pulse morphology. 
The observed double peak structure in EGRET bands rules out 
 nearly aligned rotator.  Assuming that the radio beam from 
the Geminga pulsar is oriented in a different direction from the 
 line of sight, a greatly inclined magnetic axis is also ruled out. 
 The observed phase separation of two peaks in 
$\gamma$-ray pulse profiles is explained by  
a viewing angle $\xi \sim 90^{\circ}$. 
We conclude that 
the Geminga pulsar has a inclination angle that $\alpha\sim 50^{\circ}$ and 
a viewing angle that $\xi\sim 90^{\circ}$.

4.~We applied our method to another four middle-aged radio  pulsars,
 whose spin-down power and distant from  the
 Earth  expect the  potential of the detection of the 
$\gamma$-ray emissions from those pulsars by Fermi telescope. Applying the
 inclination angle and the viewing angle inferred from the radio polarization
 characteristics, we discussed the visibility of the $\gamma$-ray emissions 
from the outer gap. We predict that $\gamma$-ray emissions from 
 PSRs B0355+54, B1449-64 and B1929+10 are detectable by Fermi telescope. 
For PSR B0740-28, the $\gamma$-ray beam from the outer gap 
will be oriented in a different direction from the viewing angle.
The observations of the PSR B0740-28 by  Fermi telescope will probably 
 be useful  for  
studying  the emission site of not only the $\gamma$-rays
 but also the radio wave in the pulsar magnetospheres. 

\section*{Acknowledgments}

The authors appreciate fruitful discussion with  
K.S. Cheng, K.Hirotani, K.S. Cheng, S. Shibata, and R.Taam. 
Authors also thank  anonymous
referee for  his/her insightful
comments on the manuscript. This work was supported by the Theoretical
Institute for Advanced Research in Astrophysics (TIARA) operated under
Academia Sinica and National Science Council Excellence Projects
program in Taiwan administered through grant number NSC 96-2752-M-007-002-PAE.

\label{lastpage}


\begin{thebibliography}{99}
\bibitem[\protect\citeauthoryear{Arons}{1983}]{ar83}
Arons J., 1983, ApJ, 266, 215
\bibitem[\protect\citeauthoryear{Becker}{2006}]{Be06}
Becker W., et al., 2006,  ApJ, 645, 1421
\bibitem[\protect\citeauthoryear{Bonsignori-Facondi}{1973}]{bo73}
Bonsignori-Facondi S.R., Salter C.J. \&  Sutton J.M., 1973, A\&A, 27, 67
\bibitem[\protect\citeauthoryear{Chang}{2007}]{ch07}
 Chang H.-K., Boggs S., Chang Y.-H. for the NCT collaboration, 
2007, AdSpR, 40, 1281
\bibitem[\protect\citeauthoryear{Cheng}{1986a}]{ch86a}
 Cheng K.S., Ho C. \&  Ruderman M. 1986a, ApJ, 300, 500 
\bibitem[\protect\citeauthoryear{Cheng}{1986b}]{ch86b}
 Cheng K.S., Ho C. \& Ruderman M. 1986b, ApJ, 300, 522
\bibitem[\protect\citeauthoryear{Cheng}{1996}]{ch96}
Cheng K.S., \& Zhang L. 1996, ApJ, 463, 271
\bibitem[\protect\citeauthoryear{Cheng}{2000}]{ch00}
  Cheng K.S., Ruderman M. \& Zhang L. 2000, ApJ, 537, 964
\bibitem[\protect\citeauthoryear{Contopoulos}{1999}]{co00}
 Contopoulos I., Kazanas D. \& Fendt C. 1999, ApJ, 511, 351
\bibitem[\protect\citeauthoryear{Daugherty}{1996}]{da96}
Daugherty J.K. \&  Harding, A.K., 1996, ApJ, 458, 278

\bibitem[\protect\citeauthoryear{Dean}{2008}]{de08}
Dean et al., 2008, Sci, 321, 1183
\bibitem[\protect\citeauthoryear{Dyks}{2003}]{dy03}
Dyks J. \& Rudak B.,  2003, ApJ, 598, 1201
\bibitem[\protect\citeauthoryear{Dyks}{2004}]{dy04}
Dyks J., Rudak B. \& Harding A.K.,  2004, ApJ, 607, 939
\bibitem[\protect\citeauthoryear{Everett}{2001}]{ev01}	
Everett J.E.\& Weisberg J.M., 2001, ApJ, 553, 341
\bibitem[\protect\citeauthoryear{Fierro}{1998}]{fi98}
Fierro J.M., Michelson P.F., Nolan P.L. \&  Thompson D.J., 1998, ApJ,
494, 734
\bibitem[\protect\citeauthoryear{Gruzinov}{2005}]{gr05}
Grizov A., 2005, Phys. Rev. Lett. 94, 021101
\bibitem[\protect\citeauthoryear{Harding}{2008}]{ha08}	
Harding A.K., Stern J.V., Dyks J. \&  Frackowiak M., 2008, ApJ, 680, 1378
\bibitem[\protect\citeauthoryear{Harding}{2002}]{ha02}
Harding A.K., Strickman M.S., Gwinn C, Dodson R., Moffet D. \& 
 McCulloch P., 2002, ApJ, 576, 376
\bibitem[\protect\citeauthoryear{Harding}{2005}]{ha05}
Harding A.K., Usov V.V., Muslimov A.G., 2005, ApJ, 622, 531
\bibitem[\protect\citeauthoryear{Hirotani}{2006}]{hi06}
Hirotani K., 2006, ApJ, 652, 1475 
\bibitem[\protect\citeauthoryear{Hirotani}{2007}]{hi07}
Hirotani K., 2007, ApJ, 662, 1173 
\bibitem[\protect\citeauthoryear{Hobbs}{2004}]{ho04}
Hobbs G., Lyne A.G., Kramer M., Martin C.E. \&  Jordan C., 
2004, MNRAS, 353, 1311 
\bibitem[\protect\citeauthoryear{Hui}{2008}]{hu08}
Hui C.Y. \&  Becker W.,  2008, A\&A, 486, 485
\bibitem[\protect\citeauthoryear{Jackson}{2002}]{ja02}
Jackson M.S., Halpern J.P., Gotthelf E.V. \&  Mattox J.R., 2002, ApJ, 578, 935
\bibitem[\protect\citeauthoryear{Jones}{2002}]{jo02}
Jones D.H., Stappers B.W.\& Gaensler B.M., 2002, A\&A, 389L, 1
\bibitem[\protect\citeauthoryear{Kanbach}{2005}]{ka05} 
Kanbach G., S\l owikoska A., Kellner S. \& Steinle H., 2005,2005,
 in Bulik T., Rudak B., Madejski G., eds, AIP Conf. Proc. Vol. 801, 
Astrophysical Sources of High Energy Particles and Radiation. 
American Institute of Physics, New York, p. 306
Proceeding, 801, 306
\bibitem[\protect\citeauthoryear{Kargaltsev}{2005}]{kar05}
Kargaltsev O.Y., Pavlov G.G., Zavlin V.E. \&  Romani R.W., 
2005, ApJ, 625, 307
\bibitem[\protect\citeauthoryear{Kataoka}{2005}]{kat05} 
Kataoka J., et al., 2005, Proc. SPIE, 5898, 133
\bibitem[\protect\citeauthoryear{Kijak}{2003}]{ki03} 
Kijak J.\& Gil J. 2003, A\&A, 397, 969
\bibitem[\protect\citeauthoryear{Large}{1969}]{la69}
Large M.I., Vaughan A.E. \&  Wielebinski R., 1969, Natur, 223, 1249
\bibitem[\protect\citeauthoryear{Lin}{2008}]{li08}
Lin L.C.-C. \& Chang H.-K., 2008, MNRAS, 387, 729
\bibitem[\protect\citeauthoryear{Manchester}{2005}]{ma05}
Manchester R.N., Hobbs G.B., Teoh A., Hobbs M., 2005, AJ, 129, 1993
\bibitem[\protect\citeauthoryear{Manchester}{1972}]{ma72}
Manchester R.N., Taylor J.H.\& Huguenin G.R., 1972, Nat., 240, 74
\bibitem[\protect\citeauthoryear{McGowan}{2006}]{mc06}
McGowan K.E., Vestrand W.T., Kennea J.A., Zane S.,
 Cropper M.\& C\'{o}rdova F.A., 2006, ApJ, 647, 1300 

\bibitem[\protect\citeauthoryear{Mestel}{1999}]{me99}
Mestel L., 1999, Stellar Magnetism, International Series of Monographs of 
\bibitem[\protect\citeauthoryear{Muslimov}{2003}]{mu03}
 Muslimov A.G. \& Harding A.K. 2003, ApJ, 588, 430
\bibitem[\protect\citeauthoryear{Muslimov}{2004}]{mu04}
 Muslimov A.G. \& Harding A.K. 2004, ApJ, 606, 1143
\bibitem[\protect\citeauthoryear{Muslimov}{2005}]{mu05}
 Muslimov A.G. \& Harding A.K. 2005, ApJ, 630, 454

\bibitem[\protect\citeauthoryear{Rankin}{1993}]{ra93}
Rankin J.M., 1993, ApJS, 84, 145
\bibitem[\protect\citeauthoryear{Romani}{1995}]{ro95}
  Romani R.W. \& Yadigaroglu I.-A. 1995, ApJ, 438, 314
\bibitem[\protect\citeauthoryear{Romani}{1996}]{ro96}
  Romani R.W. 1996, ApJ, 470, 469
\bibitem[\protect\citeauthoryear{Ruderman}{1975}]{ru75}
Ruderman M.A. \& Sutherland P.G., 1975, ApJ, 196, 51
\bibitem[\protect\citeauthoryear{Senecha}{1995}]{se95}	
Senecha V.K., Bhat C.L., Rawat H.S., Rannot R.C., Sapru M.L.,
 Kaul R.K. \&  Tickoo A.K., 1995, A\&A, 302, 133
\bibitem[\protect\citeauthoryear{Shibata}{1995}]{sh95}
Shibata S., 1995, MNRAS, 276, 537

\bibitem[\protect\citeauthoryear{Siegman}{1993}]{si93}
Siegman B.C., Manchester R.N.\& Durdin J.M., 1993, MNRAS, 262, 449
\bibitem[\protect\citeauthoryear{Tang}{2008}]{tan08}
Tang P.S. Anisia, Takata J., Jia, J.J., Cheng K.S., 2008, ApJ, 676, 562
\bibitem[\protect\citeauthoryear{Takata}{2004}]{ta04}
Takata J., Shibata S. \&  Hirotani K. 2004, MNRAS, 354, 1120
\bibitem[\protect\citeauthoryear{Takata}{2006}]{ta06}
Takata J., Shibata S., Hirotani K. \& Chang H.-K. 2006, MNRAS, 366, 1310
\bibitem[\protect\citeauthoryear{Takata}{2007}]{ta07b} 
Takata J. \& Chang H.-K., 2007 ApJ, 670, 677
\bibitem[\protect\citeauthoryear{Takata}{2007}]{ta07a} 
Takata J., Chang H.-K. \& Cheng K.S., 2007 ApJ, 656, 1044
\bibitem[\protect\citeauthoryear{Takata}{2008}]{ta08} 
Takata J., Chang H.-K. \& Shibata S.,  2008, MNRAS, 386, 748
\bibitem[\protect\citeauthoryear{Thompson}{2003}]{th02} 
Thompson D.J., 2004, in Cheng K.S., Romero G.E., eds, Cosmic Gamma Ray
Sources. Dordrecht, Kluwer, p. 149
\bibitem[\protect\citeauthoryear{Vats}{1999}]{va99} 
Vats H.O., Singal A. K., Deshpande M.R., Iyer K.N., 
Oza R., Shah C.R. \&  Doshi S., 1999, MNRAS, 302, L65


\end{thebibliography}
\end{document}